\documentclass[a4paper]{raa}            
\usepackage{CJK}
\usepackage{graphicx,times}             
\usepackage{natbib}
\usepackage{amssymb,amsmath}
\bibpunct{(}{)}{;}{a}{}{,}

\usepackage[a4paper=true,pagebackref=true]{hyperref}
\hypersetup{colorlinks = true, linkcolor = blue, anchorcolor = red, citecolor = blue, filecolor = red, pagecolor = red, urlcolor = red}

\def\l2{{\rm L2}}

\begin{document}
\begin{CJK*}{UTF8}{gbsn}

   \title{Searching for multiple populations in star clusters using the 
China Space Station Telescope}

   \volnopage{Vol.0 (20xx) No.0, 000--000}      
   \setcounter{page}{1}          

   \author{Chengyuan Li (李程远)
      \inst{1, 2}
   \and Zhenya Zheng (郑振亚)
      \inst{3, 4}
    \and Xiaodong Li (李霄栋)
      \inst{1, 2}
     \and Xiaoying Pang (庞晓莹)
      \inst{5, 6}
     \and Baitian Tang (汤柏添)
     \inst{1, 2}
    \and Antonino P. Milone
    \inst{7, 8}
    \and Yue Wang (王悦)
    \inst{9}
    \and Haifeng Wang (王海峰)
    \inst{10}
    \and Dengkai Jiang (姜登凯)
    \inst{11,12,13}
   }

   \institute{
    School of Physics and Astronomy, Sun Yat-sen University, Daxue Road, Zhuhai, Guangdong, 519082,P.R. China; {\it lichengy5@mail.sysu.edu.cn} \\
    \and
    CSST Science Center for the Guangdong-Hong Kong-Macau Greater Bay Area, Zhuhai, 519082, P.R. China; \\
        \and
             CAS Key Laboratory for Research in Galaxies and Cosmology, Shanghai Astronomical Observatory,
Shanghai, 200030, China; \\
\and
             Division of Optical Astronomical Technologies, Shanghai Astronomical Observatory, Shanghai, 200030, P.R. China; \\
 \and
             Department of Physics, Xi'an Jiaotong-Liverpool University, 111 Ren'ai Road, Suzhou, 215123, Jiangsu, P.R. China; \\
 \and
             Shanghai Key Laboratory for Astrophysics, Shanghai Normal University, 100 Guilin Road, Shanghai, 200234, P.R. China; \\
\and
             Dipartimento di Fisica e Astronomia ``Galileo Galilei'', Universit\`{a} di Padova, Vicolo dell'Osservatorio 3, I-35122, Padua, Italy \\
\and
             Istituto Nazionale di Astrofisica - Osservatorio Astronomico di Padova, Vicolo dell'Osservatorio 5, IT-35122, Padua, Italy \\
\and
	Key Laboratory of Optical Astronomy, National Astronomical Observatories, Chinese Academy of Sciences, Beijing 100101,P.R. China \\
\and
	GEPI, Observatoire de Paris, Universit\'e PSL, CNRS, Place Jules Janssen 92195, Meudon, France \\
\and
	Yunnan Observatories, Chinese Academy of Sciences, 396 Yangfangwang, Guandu District, Kunming 650216, P.R. China \\
\and
	Key Laboratory for the Structure and Evolution of Celestial Objects, Chinese Academy of Sciences, Kunming 650216, P.R. China \\
\and
	Center for Astronomical Mega-Science, Chinese Academy of Sciences, Beijing 100012, P.R. China \\          
\vs\no
   {\small Received~~20xx month day; accepted~~20xx~~month day}}

\abstract{Multiple stellar populations (MPs) in most star clusters older than 2 Gyr, as seen by lots of spectroscopic and photometric studies, 
have led to a significant challenge to the traditional view of star formation. In this field, space-based instruments, in particular the 
{\sl Hubble Space Telescope (HST)}, have made a breakthrough as they significantly improved the efficiency of detecting MPs in crowding 
stellar fields by images. The China Space Station Telescope ({\sl CSST}) and the {\sl HST} are sensitive to a similar wavelength interval, 
but it covers a field of view which is about 5-8 times wider than that of {\sl HST}. One of its instruments, 
the Multi-Channel Imager (MCI), will have multiple filters covering a wide wavelength range from NUV to NIR, making the 
{\sl CSST} a potentially powerful tool for studying MPs in clusters. In this work, we evaluate the efficiency of the designed filters 
for the MCI/{\sl CSST} in revealing MPs in different color-magnitude diagrams (CMDs). We find that CMDs made with MCI/{\sl CSST} photometry in appropriate UV filters are powerful tools to disentangle stellar populations with different abundances of He, C, N, O and Mg. On the contrary, the traditional CMDs are blind to multiple populations in globular clusters (GCs). We show that CSST has the potential of being the spearhead instrument for investigating MPs in GCs in the next decades.
}

   \authorrunning{Chengyuan Li et al. }            
   \titlerunning{MPs with the MCI/{\sl CSST}}  

   \maketitle

%
%
\section{Introduction} \label{S1}

About decades ago, star clusters were thought of as simple-stellar populations (SSPs), i.e., a sample of stars with different masses 
that are identical in age and metallicity. The traditional star formation picture concludes that all stars formed in clustered environments 
should inherit the same chemical composition of their parental molecular cloud, thus are chemically homogeneous, and are coeval 
because the strong initial stellar feedback makes the star formation mode a burst \citep[e.g.,][]{Calu15a}. This view is discarded because of 
the detection of star-to-star chemical variations in almost all globular clusters \citep[see the review of][]{Grat19a}, known as 
multiple stellar populations (MPs).


Significant chemical variations in star clusters can be detected among many light elements. Some of the most common elements 
include He, C, N, O, Na, Mg, Al.  { Extensive studies have shown that GCs contain more than one stellar population at different 
evolutionary stages}: main-sequence \citep[MS,][]{Piot07a,Milo19a}, red-giant branch \citep[RGB,][]{Carr03a,Mucc15a,Milo17a,Lato19a,Milo20a}, horizontal branch \citep[HB,][]{Grat11a},  and asymptotic giant branch \citep[AGB,][]{Wang16a,Mari17a}. MP phenomenon is likely a global feature for stars at different stages, starting from the MS to evolved giants \citep[e.g., 47 Tuc, NGC 1851,][]{Milo12a,Cumm14a,Gruy17a,Yong15a}. In 
this article, we do not discuss iron-complex clusters \citep[Type II GCs, e.g.,][]{Mari15a,Mari19a,Mari21a}. These clusters comprise the 
most massive GCs in the Galaxy \citep[e.g., $\omega$ Cen,][]{John10a}. Their origins may differ from most mono-metallic 
globular clusters (GCs), such as a tidally disrupted dwarf galaxy.

The elemental abundance variations are correlated with each other. Observations show that for clusters with MPs, the total abundance of 
C, N, O usually remains unchanged, $\delta{\rm [(C+N+O)/Fe]}\sim0$ \citep{Cohe05a,Mari16a}. Since most GCs have constant overall CNO abundance, an increase in nitrogen corresponds to depletion in carbon and oxygen. Moreover, in many metal poor GCs with MPs, sodium 
anti-correlates with oxygen \citep[e.g.,][]{Carr06a,Carr09a,Carr09b}, The abundances of magnesium and aluminum are anti-correlated as well, 
but this pattern disappears in metal-rich GCs \citep[or very weak,][]{Panc17a}.

The MP phenomenon is very common in old GCs, and it exhibits links to global parameters. A lots of comprehensive analyses has shown that the clusters total mass is a key parameter which controls some properties of MPs \citep[e.g., the fraction of 2G stars and the internal helium and nitrogen variations,][]{Milo17a,Milo18a}. { Indeed, many scenarios proposed to account for the origin of MPs strongly indicate the importance of cluster total masses.} In these scenarios, the polluters include intermediate-age AGB stars \citep{Derc10a}, fast-rotating massive dwarfs \citep{Decr07a}, massive interacting binaries \citep{deMi09a}, and single supermassive MS stars \citep{Deni14a}. Studies of clusters in the Magellanic clouds show that some clusters older than $\sim$2 Gyr have hint of MPs, while their younger counterparts have not \citep{Mart18a,Milo20a}. This has led to a hypothesis that cluster age determines the occurrence of MPs, perhaps linked to some non-standard stellar evolutionary processes \citep[][however, see \cite{Li20a,Li21a}]{Bast18a}. However, so far only a few YMCs (younger than $\sim$ 6 Gyr) have been scrutinized in terms of their stellar populations \citep[][their Fig.7]{Grat19a}. It remains unclear if the phenomenon of MPs is a common pattern for young massive clusters (YMCs), which may represent the infant stage of GCs we see today.

{ Both spectroscopic and photometric studies provide complementary information for understanding the MPs. Spectroscopic studies have provided many details about the chemical properties of stellar populations. Mostly, this method is sensitive to the brightest stars. The UV-optical-based photometry is a high efficiency method that overcomes the effect of crowding of clusters, allowing us to examine millions of stars at different evolutionary stages and positions, even in the densest regions of clusters. This is important because to fully understand the origin of MPs, a large sample of clusters covering an extensive parameter space is required. An example of photometric studies of MPs is the breakthrough made by the {\sl Hubble Space Telescope} ({\sl HST}).} The high spatial resolution of the {\sl HST} allows deep observations into clusters' core region, resulting in clean color-magnitude diagrams with subtle features. Actually, the effect of the He variation in GCs is detected through photometry rather than spectroscopy \citep[see][]{Bell10a,Piot07a} for early helium determination in $\omega$ Cen and NGC 2808). Because the helium line associated with photospheric transitions can be detected only in stars hotter than 8,500 K. Moreover, the abundances inferred from HB stars hotter than $\sim$11,500 K are not representative of the stellar helium content. As a consequence, helium abundance can be only estimated in HB stars that span a small temperature interval\footnote{The He II 10830 \AA $\,$ absorption line can be used for He abundance determination, but it requires expensive high-resolution spectra in infrared passband.} \citep[e.g.,][]{Mari14a}. 

Because He enrichment will reduce the stellar atmospheric opacity and 
increase the interior mean molecular weight, stars with different He abundances will have different surface temperatures and nuclear 
burning rates in each stage, thus complicating the CMD \citep[e.g.,][]{Piot07a,Milo15a}. For other light elements, such as C, N, O, their variations can be seen in filters 
encompassing wavelengths shorter than $\sim4000$ \AA $\;$in late-type stars. Late-type stars with same global parameters 
($T_{\rm eff}$, $\log{g}$, [Fe/H]) but are different in CNO abundances will have their spectral energy 
distribution (SED) almost identical in optical passband but different in the ultraviolet (UV) \citep[e.g.,][]{Milo15b,Milo15c}. Because most CNO-related features are distributed in 
the wavelength range of UV to blue: as examples, most O-absorptions dominate the range of $\lambda\leq3000$ \AA, the NH-, CN- and CH-absorptions are centered at $\lambda=3370$ \AA,\;3883 \AA\, and 4300 \AA, respectively. In addition, the Mg II doublet centered at 2795/2805 \AA\, is one of the most important UV-absorption features, which can be detected by corresponding UV-passbands \citep[e.g.,][]{Milo20b}. We recommend \cite{Milo20} as a nice summary. Indeed, as shown by the {\sl HST UV Legacy Survey of Galactic Globular Clusters}, almost all Galactic GCs contain MPs \citep{Piot15a,Milo17a}, undoubtedly indicating that UV-optical telescopes are powerful tools for studying MPs.

The China Space Station Telescope ({\sl CSST}) is a two-meter UV-optical space telescope, co-orbiting with the China Manned Space Station, which will be launched around $\sim$2024. The Multi-Channel Imager (MCI), a three-channel simultaneous imaging covering a wavelength range of 
0.255--1 $\mu$m, is one of the five instruments on board the {\sl CSST}. The MCI has a spatial resolution similar to the {\sl HST} ($0.18''$ at 633 nm). Its field of view is $7.5'\times7.5'$, about 500\% of the ACS/WFC@{\sl HST} ($202''\times202''$) and 770\% of the UVIS/WFC3@{\sl HST} ($162''\times162''$), which thus covers larger fields of star clusters than the {\sl HST}. It will be equipped with a 9K$\times$9K CCD, and 30 filters with different bandwidths (10 per channel). Overall, CSST is an {\sl HST}-like next generation space telescope with similar wavelength coverage and spatial resolution and a larger FoV. For details about the {\sl CSST}, we refer to the review of \cite{Zhan21a}. 

It is, therefore, essential to study the effect of MPs in the MCI/{\sl CSST} photometric system. This work aims to find the most suitable filters which can maximize the color separation between MPs in CMDs. The article is organized as follows, in Section~\ref{S2}, we introduce the details of our method. We show our main results in Section~\ref{S3}. A brief discussion of our results is present in Section~\ref{S4}.


\section{Bolometric corrections for multiple stellar populations}\label{S2}
\begin{figure}
    \centering
    \includegraphics[width=0.9\textwidth]{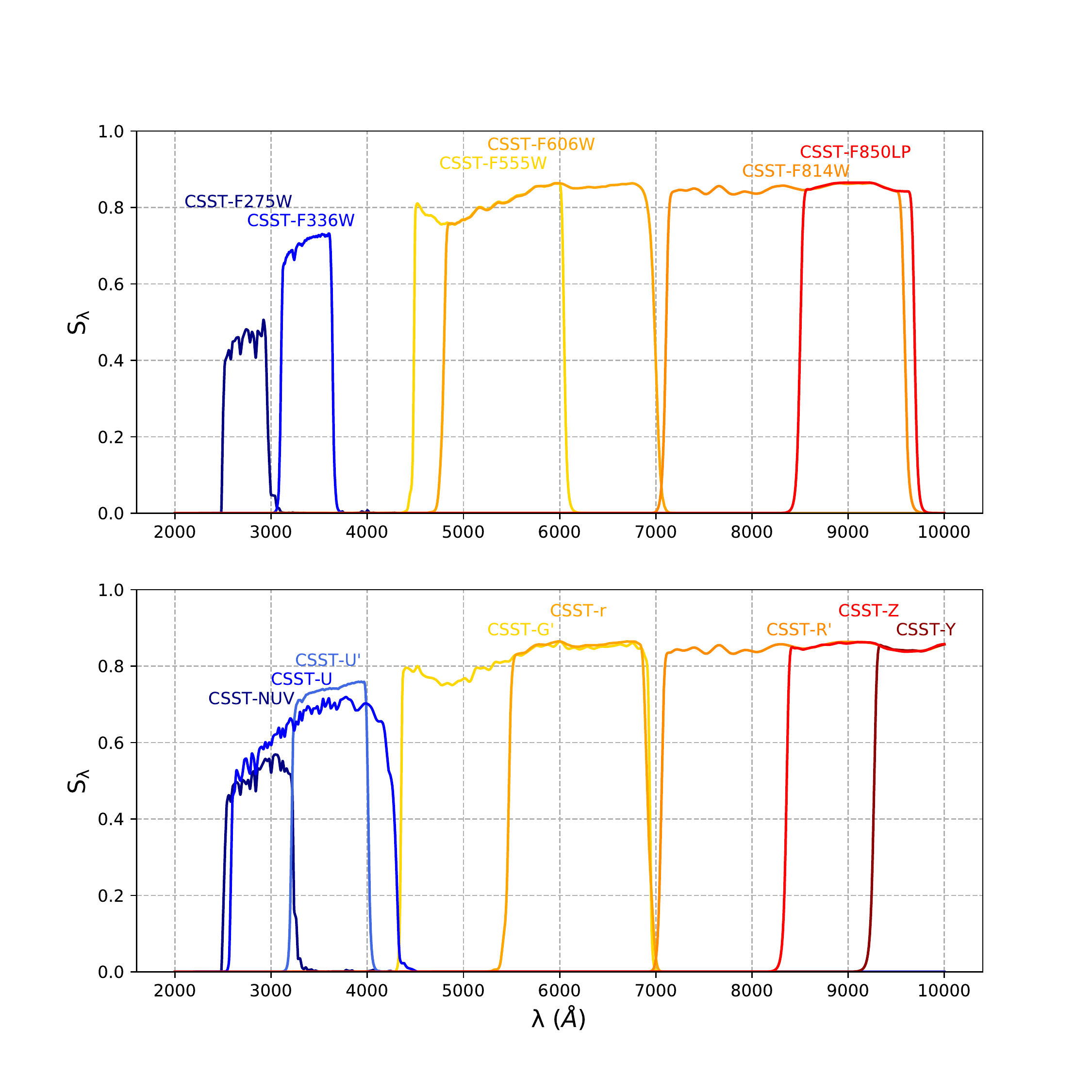}
    \caption{The total transmission curves for MCI/{\sl CSST} wide filter bands, with detector quantum efficiency being considered.}
    \label{f1}
\end{figure}

\begin{figure}
    \centering
    \includegraphics[width=0.9\textwidth]{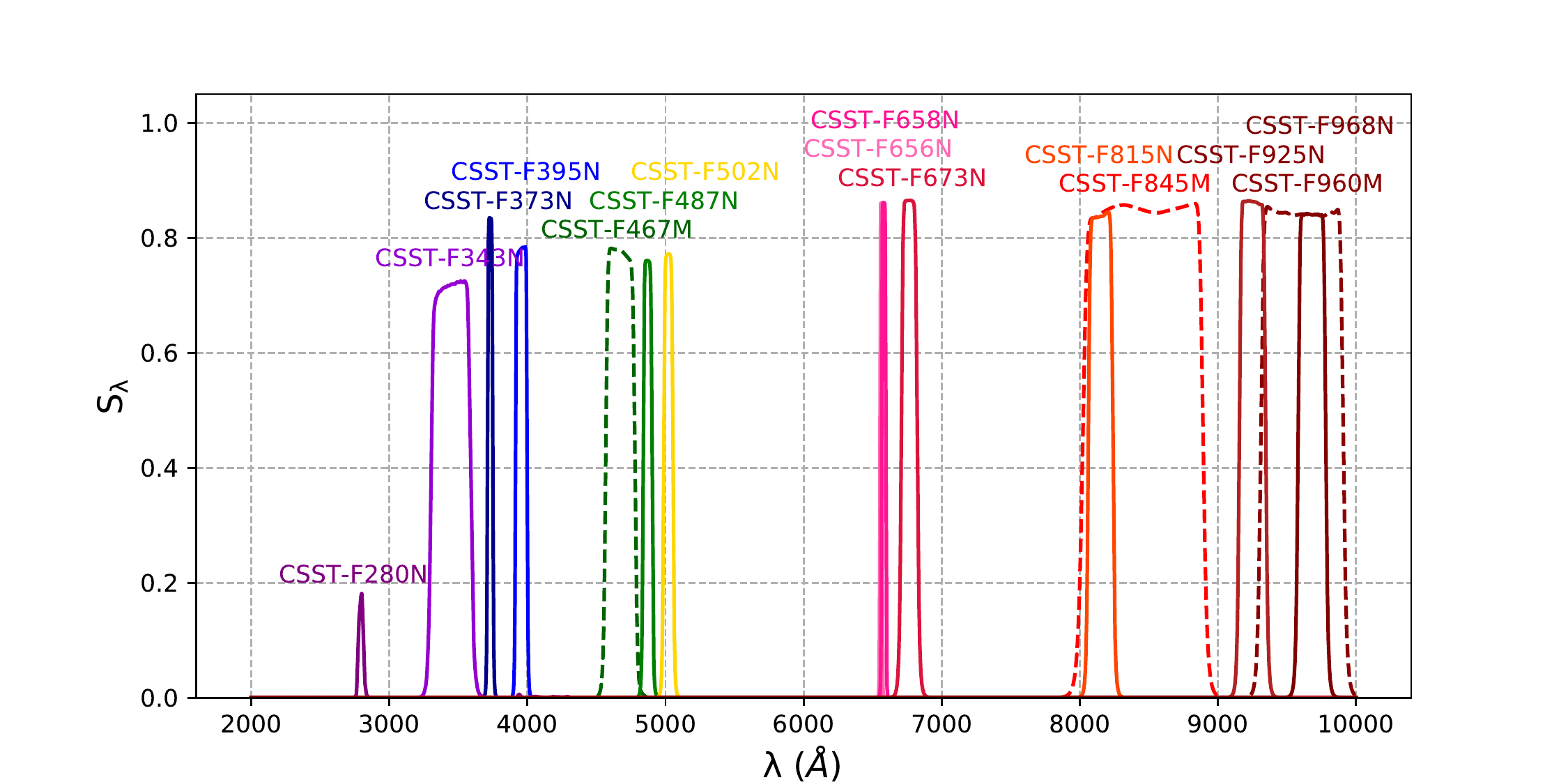}
    \caption{As fig.\ref{f1}, for narrow and medium (dashed curves) filters of the MCI/{\sl CSST}.}
    \label{f2}
\end{figure}

The current designation of the MCI/{\sl CSST} photometric system includes 15 wide passbands (FWHM/$\lambda_{\rm c}>$15\%, where $\lambda_{\rm c}$ is the mean wavelength of the passband), 3 median passbands ($5\%<{\rm FWHM}/\lambda_{\rm c}<$10\%) and 12 narrow passbands (${\rm FWHM}/\lambda_{\rm c}<$3\%, except for the {\sf CSST-f343n} filter, which follows the definition of the {\sl HST}/WFC3-F343N). 
There are two dichroic filters in the light path of MCI, which divide the wavelength range of MCI into three channels at 255 nm--430 nm, 430 nm--700 nm, and 700 nm--1000 nm. These dichroic filters prevent the use of filters such as $g$ and $i$ bands. In Table \ref{T1} we summarize some basic information of the adopted filters, including (1), their mean wavelengths, $\lambda_{\rm c}$; (2), the full-width-half-maximum, the FWHM; (3) the corresponding wavelengths of 50\% of the maximum transmission curve (left: $\lambda_{\rm L50}$, right: $\lambda_{\rm R50}$); (4) the steepness of the transmission curve, describing by Tan$_{x}$=$\Delta{\lambda_x}/\lambda_x$, where $x$ are L50/R50 and $\Delta{\lambda_x}$ is the wavelength difference between 0\% and 100\% maximum transmission, and (5), the average transmission efficiency within FWHM, T50. The transmission curves for MCI/CSST filters are present in Fig.\ref{f1} (wide passbands) and Fig.\ref{f2} (narrow and medium passbands). We do not show the transmission curve for the {\sf CSST-WU} passband because it is being redesigned. 

\begin{table*}
\caption{MCI/{\sl CSST} filter parameters (ranked by increasing mean wavelengths)
}
\begin{center}
\begin{tabular}{lccccccc}
\hline
%
%
Filter Names & $\lambda_{\rm c}$ & FWHM & $\lambda_{\rm L50}$ & $\lambda_{\rm R50}$ & Tan$_{\rm L50}$ & Tan$_{\rm R50}$ & T50 \\
  & (nm) & (nm) & (nm) & (nm) & & & \\
\hline
\multicolumn{8}{c}{Channel 1 (UV)}  
\\\hline
{\sf CSST-f275w} & 272.5 & 45.0 & 250.0$\pm$2.5 & 295.0$\pm$2.5 & 0.05 & 0.05 & 60\% \\
{\sf CSST-f280n} & 279.6 & 4.1 & 277.5$\pm$0.4 & 281.6$\pm$0.4 & 0.02 & 0.02 & 25\% \\
{\sf CSST-NUV} & 286.5 & 69.1 & 251.9$\pm$2.5 & 321.0$\pm$2.5 & 0.05 & 0.02 & 65\% \\
{\sf CSST-f336w} & 337.1 & 54.0 & 310.1$\pm$3.0 & 364.1$\pm$3.0 & 0.03 & 0.03 & 80\% \\
{\sf CSST-U$'$} & 340.0 & 170.0 & 255.0$\pm$3.0 & 425.0$\pm$3.0 & 0.08 & 0.05 & 80\% \\
{\sf CSST-f343n} & 344.5 & 28.9 & 330.0$\pm$2.0 & 358.9$\pm$2.0 & 0.02 & 0.02 & 75\% \\
{\sf CSST-u} & 361.2 & 80.4 & 321.0$\pm$2.5 & 401.4$\pm$2.5 & 0.02 & 0.02 & 80\% \\
{\sf CSST-f373n} & 373.0 & 4.9 & 370.5$\pm$0.5 & 375.4$\pm$0.5 & 0.01 & 0.01 & 75\% \\
{\sf CSST-f395n} & 395.5 & 8.4 & 391.3$\pm$0.5 & 399.7$\pm$0.5 & 0.01 & 0.01 & 85\% \\
\hline
\multicolumn{8}{c}{Channel 2 (visible)} 
\\\hline 
{\sf CSST-f467m} & 468.4 & 21.5 & 457.6$\pm$3.0 & 479.1$\pm$3.0 & 0.01 & 0.01 & 90\% \\
{\sf CSST-f487n} & 487.2 & 6.0 & 484.2$\pm$0.5 & 490.2$\pm$0.5 & 0.01 & 0.01 & 80\% \\
{\sf CSST-f502n} & 501.0 & 6.6 & 497.7$\pm$0.5 & 504.3$\pm$0.5 & 0.01 & 0.01 & 80\% \\
{\sf CSST-f555w} & 526.7 & 159.1 & 447.1$\pm$3.0 & 606.2$\pm$3.5 & 0.02 & 0.02 & 90\% \\
{\sf CSST-G$'$} & 565.0 & 260.0 & 435.0$\pm$3.0 & 695.0$\pm$4.0 & 0.02 & 0.02 & 90\% \\
{\sf CSST-f606w} & 594.7 & 229.1 & 480.1$\pm$3.0 & 709.2$\pm$4.0 & 0.02 & 0.02 & 90\% \\
{\sf CSST-r} & 619.5 & 145.3 & 546.8$\pm$3.0 & 692.1$\pm$3.5 & 0.02 & 0.02 & 90\% \\
{\sf CSST-f656n} & 656.2 & 1.7 & 655.3$\pm$0.2 & 657.0$\pm$0.2 & 0.01 & 0.01 & 80\% \\
{\sf CSST-f658n} & 658.6 & 2.7 & 657.2$\pm$0.5 & 659.9$\pm$0.5 & 0.01 & 0.01 & 80\% \\
{\sf CSST-f673n} & 676.6 & 11.9 & 670.6$\pm$1.0 & 682.5$\pm$1.0 & 0.01 & 0.01 & 85\% \\
\hline
\multicolumn{8}{c}{Channel 3 (visible--NIR)} 
\\\hline 
{\sf CSST-f815n} & 815.0 & 20.0 & 805.2$\pm$2.0 & 825.2$\pm$2.0 & 0.01 & 0.01 & 90\% \\
{\sf CSST-f814w} & 833.8 & 253.1 & 707.2$\pm$4.0 & 960.3$\pm$4.0 & 0.02 & 0.02 & 92\% \\
{\sf CSST-f845m} & 847.2 & 88.0 & 803.2$\pm$4.0 & 891.2$\pm$4.0 & 0.02 & 0.02 & 92\% \\
{\sf CSST-R$'$} & 852.5 & 295.0 & 705.0$\pm$5.0 & 1000.0 & 0.02 & - & 92\% \\
{\sf CSST-z} & 921.2 & 175.6 & 824.4$\pm$4.0 & 1000.0 & 0.02 & - & 92\% \\
{\sf CSST-f925n} & 925.0 & 30.0 & 910.0$\pm$2.0 & 940.0$\pm$2.0 & 0.01 & 0.01 & 90\% \\
{\sf CSST-f850lp} & 927.6 & 144.8 & 855.2$\pm$4.0 & 1000.0 & 0.01 & - & 92\% \\
{\sf CSST-y} & 963.4 & 72.4 & 926.7$\pm$5.0 & 1000.0 & 0.02 & - & 92\% \\
{\sf CSST-f960m} & 960.0 & 60.0 & 930.0$\pm$4.0 & 990.0$\pm$4.0 & 0.01 & 0.01 & 92\% \\
{\sf CSST-f968n} & 968.0 & 20.0 & 958.0$\pm$2.0 & 978.0$\pm$2.0 & 0.01 & 0.01 & 90\% \\
\hline 
\end{tabular}
\end{center}

\label{T1}
\end{table*}

To evaluate the effect of MPs on the MCI/{\sl CSST} photometric system, { we need to select a given chemical pattern to simulate its effect on  photometry. In this work, we studied two cases, an NGC 2808-like MPs and a less extreme case with chemical variations half of the NGC 2808. We 
study these two cases because NGC 2808 represents a quintessential (extreme) example of mono-metallic GC that exhibits variations in almost all light elements \citep{Piot07a,Grat11a,Mucc15a,Wang16a,Lato19a}. The half-NGC 2808 chemical pattern can represent many less massive GCs with MPs \citep[see,][]{Milo18a}.} { We used the Dartmouth Stellar Evolution Database (Dartmouth model) to generate isochrones with different He abundances \citep{Dott08a}. We define two stellar populations following described by two isochrones to represent an NGC 2808-like MPs.
The Dartmouth model allows us to interpolate any isochrone with parameters within $-$2.5$<$[Fe/H]$<$+0.5 dex, initial He abundances from $Y$=0.245 to 0.40, and ages from 1 Gyr to 15 Gyr. It is a suitable stellar model to describe an NGC 2808-like GC with an extreme helium enrichment.}
Based on the literature of \cite{Piot07a}, we adopt the age and metallicity of our two populations, both $t$=12 Gyr and [Fe/H]=$-$1.0 dex, 
respectively. They are different in He abundance, with $Y$=0.25 for one population and $Y$=0.40 for another (hereafter 1P and 2P). The [$\alpha$/Fe] 
is 0.0 dex as we confirm that it has a negligible effect on the derived isochrones. 

{ The derived isochrones contain a series of physical parameters, including stellar masses ($M/M_{\odot}$), effective surface temperatures ($\log{T_{\rm eff}}$), surface gravity ($\log{g}$) and luminosities ($\log{L/L_{\odot}}$). These parameters, along with the adopted metallicity ([Fe/H]), are used for calculating the bolometric corrections through the PARSEC database of bolometric correction \citep[The YBC database,][]{Chen19a}. The YBC database provides absolute magnitudes in the CSST filter set (we obtain the specific MCI/CSST magnitudes through private communication with Dr. Chen Yang).} 

{ We used the package {\sc Spectrum} (version 2.77) to calculate isochrones with specific elemental abundances, thus loci for stellar populations. Given a stellar atmosphere model and certain inputs, {\sc Spectrum} calculates a synthetic stellar spectrum with input parameters. The {\sc Spectrum} can calculate reliable synthetic B- to mid-M-type stellar spectra. The synthetic spectra are calculated under the silent \& isotope \& {\sc Atlas} modes. The input line lists are obtained from the main page of the {\sc Spectrum} package\footnote{\url{http://www.appstate.edu/~grayro/spectrum/ftp/download.html}}, which includes more than $1.6\times10^8$ absorption lines covering a wavelength range of 900\AA -- 40000\AA. The input atmosphere models were computed with the {\sc Atlas9} model atmosphere program written by \cite{Kuru93a}\footnote{\url{https://wwwuser.oats.inaf.it/castelli/grids.html}} with parameters ([Fe/H], $\log{g}$, $T_{\rm eff}$) from the base isochrone.} 

{ We calculate synthetic spectra with different elemental abundances for 1P and 2P stars}. Under the adoption of [Fe/H]=$-$1.0 dex and assuming that the total abundance of C,N,O remains unchanged, we set all 2P stars to have $\delta$[N/Fe]=$+$1.0 dex, and $\delta$[O/Fe]=$\delta$[C/Fe]=$-$0.5 dex. 
Spectroscopic studies show that NGC 2808 has a Na-O anti-correlation for its member stars \cite{Carr09a}. We adopt $\delta$[Na/Fe]=$+$0.5 dex through visual inspection \citep[the Fig.7 of][]{Carr09a}, corresponding to a 0.5 dex depletion of the O abundance. According to \cite{Panc17a}, 
NGC 2808 also exhibits Mg-Al anti-correlation, we set a $\delta$[Mg/Fe]=$-$0.5 dex and $\delta$[Al/Fe]=$+$1.0 dex for 2P stars in our model. 
We will examine the effects of an NGC 2808-like MPs both with/without a $Y$ variation (hereafter Case 1 and Case 2), { and a less extreme case with all elements variations is half of NGC 2808 (no $Y$ variation, Case 3).} We also test other effects of individual element variation, including (1), the $Y$ variation, (2), the CNO variation, (3) the Na variation, (4) the Mg variation and (5) the Al variation, as well. In these cases, we (not-so-)arbitrarily assume reasonable variations based on literatures \citep{Piot07a,Carr09a,Panc17a}. We summarize these adopted models in Table \ref{T2}. { The isochrones for chemically enriched stellar populations, thus 2P loci, are calculated as follows, 
\begin{equation}
M_i=-2.5\log{\frac{\int\nolimits_{\lambda_1}^{\lambda_2}f_{\lambda}S_{\lambda,i}d\lambda}{\int\nolimits_{\lambda_1}^{\lambda_2}f_{\lambda}^0S_{\lambda,i}d\lambda}}+M_{i,0}
\end{equation}
where $M_i$ is the expected absolute magnitude for a chemically enriched star observed through the filter band $i$ (See Table \ref{T1}), $M_{i,0}$ is the corresponding absolute magnitude for the counterpart with normal chemical abundance. $f_\lambda$ and $f_\lambda^0$ are their radiative fluxes (at 10 pc)  we received at the central wavelength of $\lambda$, respectively, which are calculated through the {\sc Spectrum}. $S_{\lambda,i}$ is the transmission curve of a specific filter band $i$. The quantities $\lambda_1$ and $\lambda_2$ indicate the lower and upper wavelength limits. According to the wavelength range of the {\sl CSST}, we set them as 2500 \AA\;and 10000 \AA, respectively. }

\begin{table*}
\caption{Adopted models with different abundance variations for 2P stars
}
\begin{center}
\begin{tabular}{lccccccc}
\hline
%
%
Model names & $\delta{Y}$ & $\delta$[C/Fe] & $\delta$[N/Fe] & $\delta$[O/Fe] & $\delta$[Na/Fe] & $\delta$[Mg/Fe] & $\delta$[Al/Fe] \\
  & & (dex) & (dex) & (dex) & (dex) & (dex) & (dex) \\
\hline
He variation & 0.15 & 0.0 & 0.0 & 0.0 & 0.0 & 0.0 & 0.0 \\
CNO variation & 0.0 & $-$0.5 & $+$1.0 & $-$0.5 & 0.0 & 0.0 & 0.0 \\
Na variation & 0.0 & 0.0 & 0.0 & 0.0 & $+$0.5 & 0.0 & 0.0 \\
Mg variation & 0.0 & 0.0 & 0.0 & 0.0 & 0.0 & $+$0.5 & 0.0 \\
Al variation & 0.0 & 0.0 & 0.0 & 0.0 & 0.0 & 0.0 & $+$0.5 \\
Case 1 & 0.0 & $-$0.5 & $+$1.0 & $-$0.5 & $+$0.5 & $-$0.5 & $+$1.0 \\
Case 2 & 0.15 & $-$0.5 & $+$1.0 & $-$0.5 & $+$0.5 & $-$0.5 & $+$1.0 \\
Case 3 & 0.0 & $-$0.18 & $+$0.7 & $-$0.18 & $+$0.2 & $-$0.2 & $+$0.7 \\
\hline 
\end{tabular}
\end{center}
\label{T2}
\end{table*}

Finally, we will compare the loci of the 1P (described by the standard isochrone) with that of the 2P. To quantify their performances of separating two populations, we will calculate their color differences refer to the filter {\sf CSST-f814w}, $M_{i}-M_{\rm f814w}$, for a referenced RGB star (below the RGB bump) with $M_{\rm f555w}$=2.0 mag and a bottom MS star with $M_{\rm f555w}$=8.0 mag. The selection of these two stages is arbitrary, which corresponds to a K1-type giant and a K4-type dwarf. We confirm that they are sufficiently cool to exhibit prominent absorption features of CNO-related molecules. In this work, a positive color difference means the chemically enriched population stars (defined in Table \ref{T2}) are bluer than normal stars. 

\section{Color-magnitude diagrams}\label{S3}
In this section we present our main results, which include CMDs of 1P and 2P loci with different chemical patterns (Table \ref{T2}). 
\subsection{Helium variation}
Helium is the direct product of H-burning. Polluted stars with chemical anomalies must be He-enriched. As introduced, member stars of GCs are too cold to produce He absorptions at UV-optical passbands. For old stellar populations like GCs, the effect of He enrichment is evolutionary. The effect of helium is reflected by the value of the average molecular weight, $\mu$ and the opacity, $\kappa_0$. Because the helium opacity is lower than the hydrogen opacity, an increase of the helium abundance at constant metallicity will increase $\mu$ and decrease $\kappa_0$, making the stellar evolutionary track have higher luminosity ($L$) and effective temperature ($T_{\rm teff}$). In addition, the increased $L$ will make stars burn their central hydrogen faster than He-normal stars. As a result, the He-enriched population will always exhibit a lower main-sequence turnoff (MSTO) because it evolves more rapidly than its coeval He-normal stellar population. 

\begin{figure}
    \centering
    \includegraphics[width=0.9\textwidth]{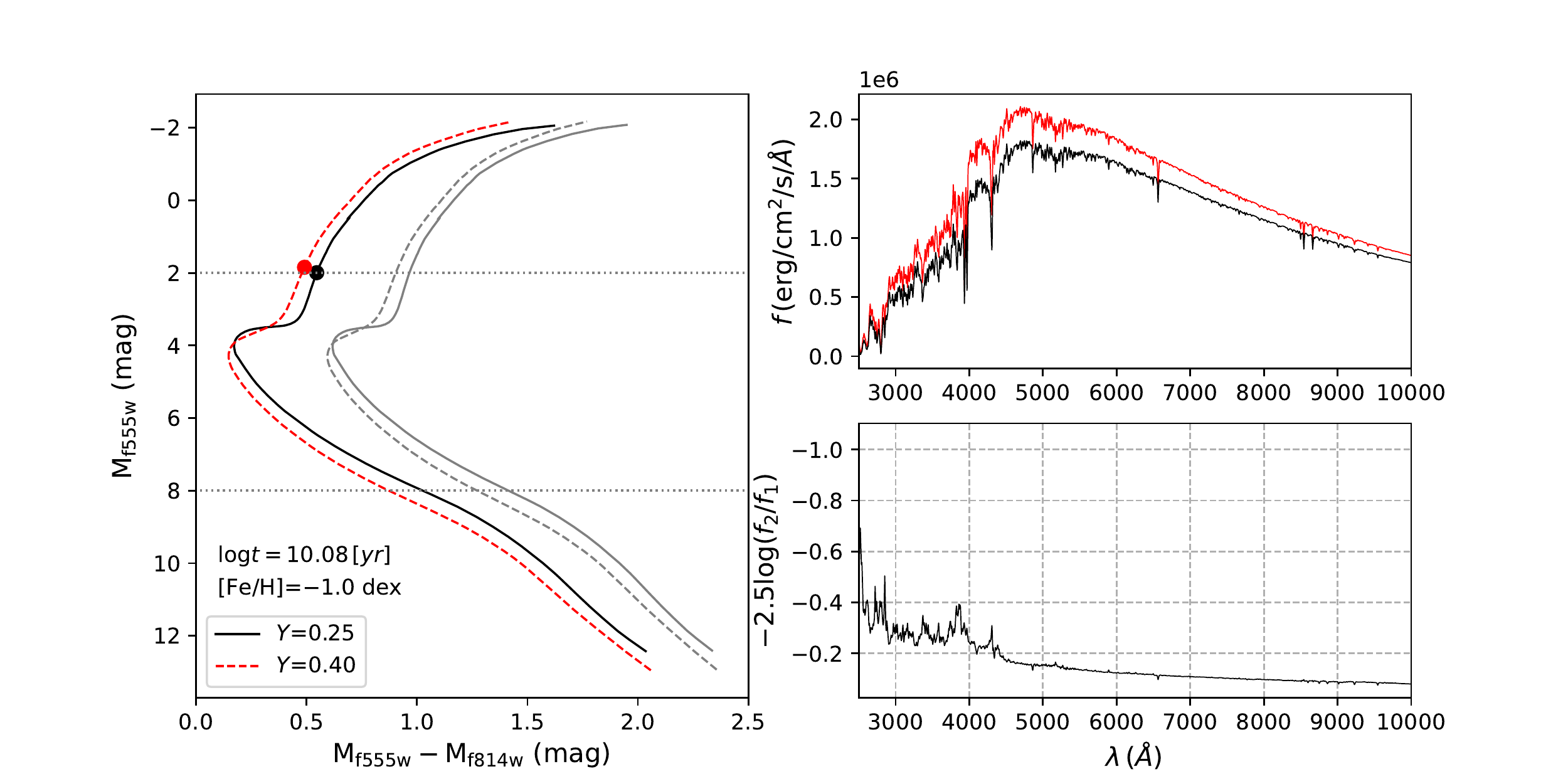}
    \caption{The effect of the Helium variation in the CMD. Left panel: the CMD in $M_{\rm f555w}-M_{\rm f814w}$ vs. $M_{\rm f555w}$ for 1P (black solid line) and 2P (He-rich, red dashed line). Grey dotted lines remarks the RGB and MS stages at $M_{\rm f555w}$=2.0 and 8.0 mag. { The same loci for 1P and 2P under the UVIS/WFC3 {\sl HST} photometric system are also present (grey solid and dashed lines).} Two example RGB stars are highlighted by black (He-normal) and red (He-rich) circles. They are at the same stage indicated by the DSEP model. The upper-right panel exhibits their synthesis spectra (black: helium-normal; red: helium-rich). The bottom-right panel shows the spectrum of their magnitude difference.}
    \label{f3}
\end{figure}

In Fig.\ref{f3} we show the isochrones for the 1P (He-normal, $Y$=0.25) and 2P ($Y$=0.40) stars (left panel). In the upper-right panel we show two giants spectra with different He abundances at the same evolutionary stages, as indicated by the black (He-normal) and red (He-rich) circles in the left panel. In the bottom-right panel we exhibit their flux ratios in terms of $-2.5\log{f_2/f_1}$ ($f_2$ and $f_1$ are fluxes of 2P and 1P stars), which is equal to a magnitude difference spectrum. Fig.\ref{f3} shows that for two stars at the same evolutionary stage, the He-rich star is brighter/bluer than the normal star. The color axis in Fig.\ref{f3} is described by $M_{\rm f555w}-M_{\rm f814w}$, for the referenced RGB(MS) stage (indicated by grey dotted lines), the color difference between two populations is $\Delta({M_{\rm f555w}-M_{\rm f814w})}\sim$0.07(0.15) mag, respectively. { As a comparison, we also show the same loci for 1P and 2P under the {\sl HST} photometric system (UVIS/WFC3), as indicated by grey solid (1P) and dashed (2P) lines in the left panel. We find that the color separations between 1P and 2P are similar in the MCI/{\sl CSST} and UVIS/WFC3 {\sl HST} filter systems.}The color differences between other passbands, $\Delta(M_{i}-M_{\rm f814w})$, are calculated, which is present in Tables \ref{T3} and \ref{T4} (for the RGB and MS stages). 

\begin{figure}
    \centering
    \includegraphics[width=0.9\textwidth]{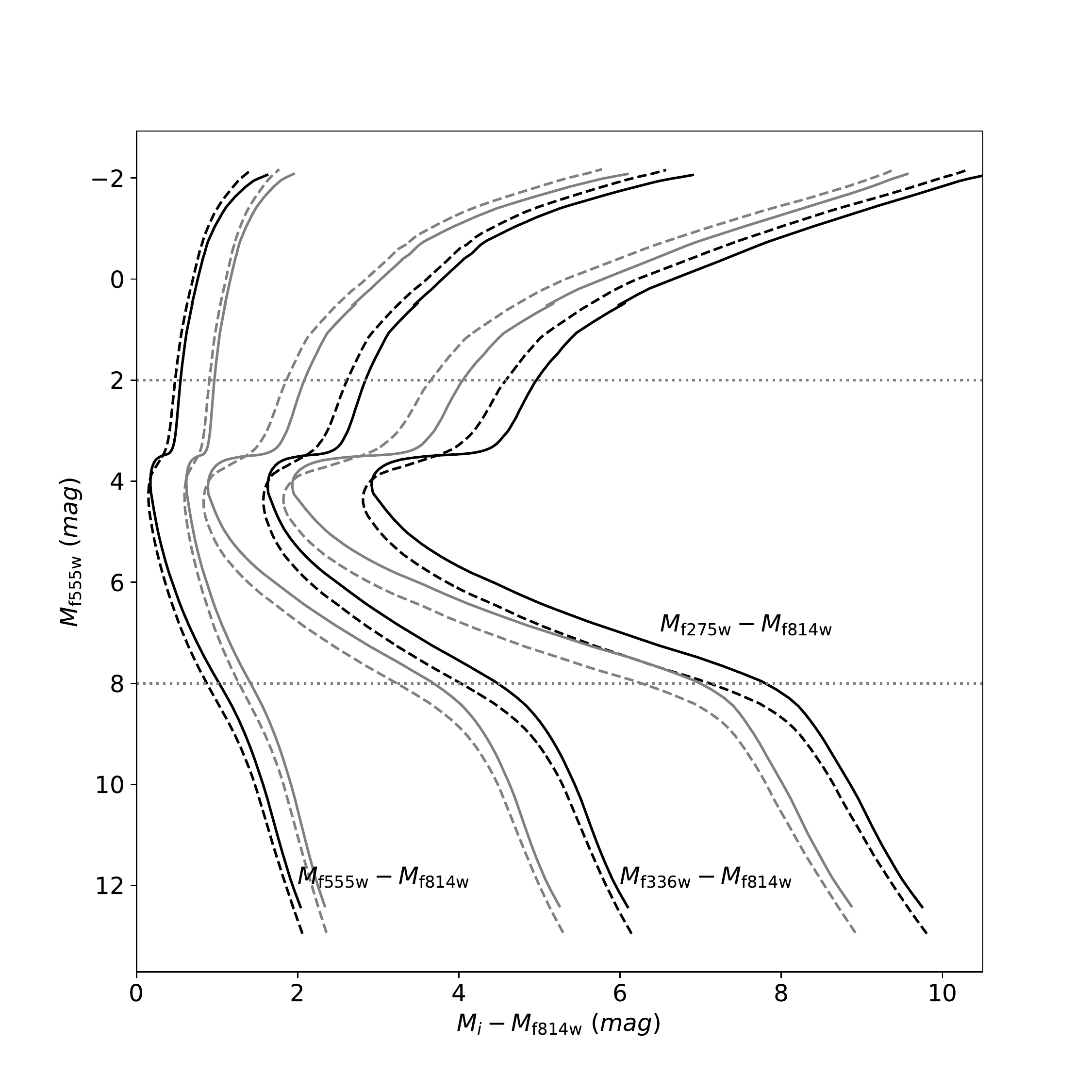}
    \caption{From left to right, loci of normal (1P, black solid lines) and helium-rich (2P, black dashed line) populations in colors of $M_{\rm f275w}-M_{\rm f814w}$, $M_{\rm f336w}-M_{\rm f814w}$ and $M_{\rm f555w}-M_{\rm f814w}$, respectively. { The same loci for 1P and 2P under the UVIS/WFC3 {\sl HST} photometric system are indicated by grey solid and dashed lines.}}
    \label{f4}
\end{figure}

\begin{figure}
    \centering
    \includegraphics[width=0.9\textwidth]{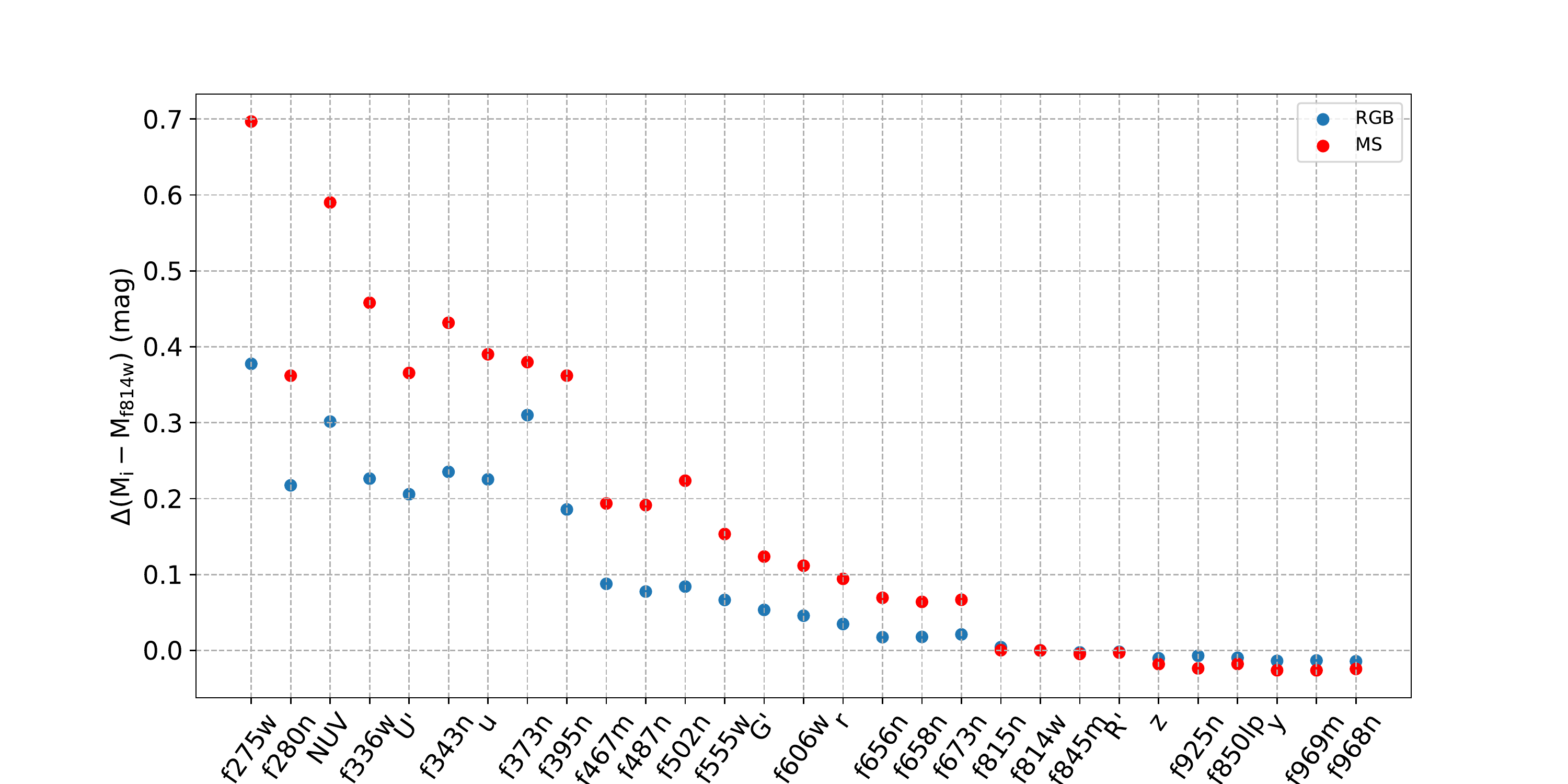}
    \caption{Color differences between normal and helium-enhanced populations in different color bands.}
    \label{f5}
\end{figure}

Fig.\ref{f3} shows that helium-rich MS and RGB stars are hotter than their normal counterparts. A wide color baseline is efficient to disentangle stellar populations with different helium abundances. In Fig.\ref{f4} we show three isochrone pairs which describe $Y$=0.25 and $Y$=0.40 populations in different colors, $M_{\rm f275w}-M_{\rm f814w}$, $M_{\rm f336w}-M_{\rm f814w}$ and $M_{\rm f555w}-M_{\rm f814w}$ (from left to right). It shows that a wider color baseline can resolve a larger color difference between the two populations. { These color differences between 1P and 2P are similar to those under the UVIS/WFC3 {\sl HST} photometric system (grey solid and dashed lines, respectively)}. Fig.\ref{f5} presents the color differences in different filter bands, $M_{\rm i}-M_{\rm f814w}$. Indeed, we find that the bluer the first filter band, the higher the color difference between the two populations. The maximum color differences at the RGB and MS stages appear both in $M_{\rm f275w}-M_{\rm f814w}$, which are $\Delta(M_{\rm f275w}-M_{\rm f814w})\sim0.4$ mag and $\sim$0.7 mag, respectively. To resolve such color differences, a minimal sign-to-noise ratio of SNR$\sim$4 for each passband is required (see discussions in Section \ref{S4}). 

\begin{table*}
\caption{Color differences of $M_{\rm i}-M_{\rm f814w}$ ($i$ indicate different filter bands, see Table \ref{T1}) between 1P and 2P (for different chemical variations [var.], see Table \ref{T2}), for the referenced RGB stage. The number associated with an asterisk means this color difference can be resolved at the distance of NGC 2808 with a total exposure time of 300 s (see discussions in Section \ref{S4}). 
}
\begin{center}
\begin{tabular}{lccccccc}
\hline
%
%
Color differences (mag) & He var. & CNO var. & Na var. & Mg var. & Al var. & Case 1 & Case 2 \\
\hline
$\Delta$($M_{\rm f275w}-M_{\rm f814w}$) & 0.378* & 0.047 & $-$0.004 & $-$0.162* & 0.012 & 0.113* & 0.489* \\
$\Delta$($M_{\rm f280n}-M_{\rm f814w}$) & 0.217* & 0.032 & $-$0.001 & $-$0.397* & 0.020 & 0.380* & 0.595* \\
$\Delta$($M_{\rm NUV}-M_{\rm f814w}$) & 0.301* & 0.075* & $-$0.003 & $-$0.092* & 0.011 & 0.104* & 0.404* \\
$\Delta$($M_{\rm f336w}-M_{\rm f814w}$) & 0.226* & $-$0.098* & $-$0.001 & $-$0.016 & 0.013 & $-$0.100* & 0.127* \\
$\Delta$($M_{\rm U'}-M_{\rm f814w}$) & 0.206* & $-$0.030* & $-$0.002 & $-$0.035* & 0.005 & $-$0.024* & 0.182* \\
$\Delta$($M_{\rm f343n}-M_{\rm f814w}$) & 0.235* & $-$0.233* & $-$0.002 & $-$0.018 & 0.014 & $-$0.236* & 0.001 \\
$\Delta$($M_{\rm u}-M_{\rm f814w}$) & 0.225* & $-$0.102* & $-$0.002 & $-$0.021* & 0.004 & $-$0.105* & 0.122* \\
$\Delta$($M_{\rm f373n}-M_{\rm f814w}$) & 0.310* & $-$0.032 & $-$0.003 & $-$0.030 & 0.017 & $-$0.037 & 0.274* \\
$\Delta$($M_{\rm f395n}-M_{\rm f814w}$) & 0.186* & $\sim$0.000 & $-$0.001 & $-$0.010 & $-$0.053 & $-$0.025 & 0.161* \\
\hline
$\Delta$($M_{\rm f467m}-M_{\rm f814w}$) & 0.088* & 0.007 & $-$0.001 & $-$0.006 & $-$0.001 & 0.007 & 0.094* \\
$\Delta$($M_{\rm f487n}-M_{\rm f814w}$) & 0.078* & 0.009 & $\sim$0.000 & $-$0.003 & 0.001 & 0.008 & 0.086* \\
$\Delta$($M_{\rm f502n}-M_{\rm f814w}$) & 0.084* & 0.006 & $-$0.001 & $-$0.010 & 0.002 & 0.006 & 0.090* \\
$\Delta$($M_{\rm f555w}-M_{\rm f814w}$) & 0.066* & 0.006* & $-$0.001 & $-$0.005* & 0.001 & 0.005* & 0.072* \\
$\Delta$($M_{\rm G'}-M_{\rm f814w}$) & 0.053* & 0.006* & $\sim$0.000 & $-$0.004* & $\sim$0.000 & 0.006* & 0.059* \\
$\Delta$($M_{\rm f606w}-M_{\rm f814w}$) & 0.004* & 0.004* & $\sim$0.000 & $-$0.003 & $\sim$0.000 & 0.004* & 0.049* \\
$\Delta$($M_{\rm r}-M_{\rm f814w}$) & 0.035* & 0.003 & $\sim$0.000 & $-$0.001 & $-$0.001 & 0.003 & 0.037* \\
$\Delta$($M_{\rm f656n}-M_{\rm f814w}$) & 0.017 & 0.002 & 0.001 & 0.012 & 0.003 & 0.003 & 0.020 \\
$\Delta$($M_{\rm f658n}-M_{\rm f814w}$) & 0.018 & 0.003 & $\sim$0.000 & 0.002 & $\sim$0.000 & 0.003 & 0.020 \\
$\Delta$($M_{\rm f673n}-M_{\rm f814w}$) & 0.021* & 0.003 & $\sim$0.000 & $-$0.001 & $-$0.001 & 0.002 & 0.023* \\
\hline
$\Delta$($M_{\rm f815n}-M_{\rm f814w}$) & 0.004 & $-$0.002 & $-$0.001 & $\sim$0.000 & $\sim$0.000 & $-$0.003 & 0.002 \\
$\Delta$($M_{\rm f845m}-M_{\rm f814w}$) & $-$0.003 & $\sim$0.000 & $\sim$0.000 & 0.001 & $\sim$0.000 & $\sim$0.000 & $-$0.002 \\
$\Delta$($M_{\rm R'}-M_{\rm f814w}$) & $-$0.002 & $\sim$0.000 & $\sim$0.000 & $\sim$0.000 & $\sim$0.000 & $\sim$0.000 & $-$0.002 \\
$\Delta$($M_{\rm z}-M_{\rm f814w}$) & $-$0.010* & $\sim$0.000 & $\sim$0.000 & $-$0.001 & $\sim$0.000 & $\sim$0.000 & $-$0.011* \\
$\Delta$($M_{\rm f925n}-M_{\rm f814w}$) & $-$0.007 & $-$0.006 & $\sim$0.000 & $-$0.001 & $\sim$0.000 & $-$0.006 & $-$0.012 \\
$\Delta$($M_{\rm f850lp}-M_{\rm f814w}$) & $-$0.010* & $-$0.001 & $\sim$0.000 & $\sim$0.000 & $\sim$0.000 & $\sim$0.000 & $-$0.010* \\
$\Delta$($M_{\rm y}-M_{\rm f814w}$) & $-$0.014* & $-$0.002 & $\sim$0.000 & $-$0.002 & $\sim$0.000 & $-$0.002 & $-$0.016* \\
$\Delta$($M_{\rm f960m}-M_{\rm f814w}$) & $-$0.013 & $-$0.002 & $\sim$0.000 & $-$0.002 & $\sim$0.000 & $-$0.002 & $-$0.015* \\
$\Delta$($M_{\rm f968n}-M_{\rm f814w}$) & $-$0.014 & $-$0.001 & $\sim$0.000 & $-$0.002 & $\sim$0.000 & $-$0.001 & $-$0.015 \\
\hline 
\end{tabular}
\end{center}
\label{T3}
\end{table*}

\begin{table*}
\caption{The same as Table \ref{T3}, except that the reference stage is the bottom MS. The number associated with a asterisk means this color difference can be resolved at the distance of NGC 2808 with a total exposure time of 54,000 s (180 exposures)}

\begin{center}
\begin{tabular}{lccccccc}
\hline
%
%
Color differences (mag) & He var. & CNO var. & Na var. & Mg var. & Al var. & Case 1 & Case 2 \\
\hline
$\Delta$($M_{\rm f275w}-M_{\rm f814w}$) & 0.697 & 0.377 & 0.019 & 0.160 & 0.012 & 0.488 & 0.918 \\
$\Delta$($M_{\rm f280n}-M_{\rm f814w}$) & 0.362 & 0.092 & 0.026 & 0.149 & 0.020 & 0.323 & 0.710 \\
$\Delta$($M_{\rm NUV}-M_{\rm f814w}$) & 0.590* & 0.327 & 0.019 & 0.209 & 0.011 & 0.352 & 0.777* \\
$\Delta$($M_{\rm f336w}-M_{\rm f814w}$) & 0.458* & $-$0.142* & 0.013 & 0.170* & 0.013 & $-$0.132* & 0.278* \\
$\Delta$($M_{\rm U'}-M_{\rm f814w}$) & 0.366* & $-$0.052* & 0.010 & 0.127* & 0.005 & $-$0.049* & 0.298* \\
$\Delta$($M_{\rm f343n}-M_{\rm f814w}$) & 0.432* & $-$0.363* & 0.013 & 0.163 & 0.014 & $-$0.329* & 0.026 \\
$\Delta$($M_{\rm u}-M_{\rm f814w}$) & 0.390* & $-$0.156* & 0.011 & 0.138* & 0.004 & $-$0.157* & 0.201* \\
$\Delta$($M_{\rm f373n}-M_{\rm f814w}$) & 0.380 & $-$0.109 & 0.016 & 0.188 & 0.017 & $-$0.073 & 0.268 \\
$\Delta$($M_{\rm f395n}-M_{\rm f814w}$) & 0.362* & $-$0.040 & 0.011 & 0.133 & $-$0.053 & $-$0.192 & 0.227 \\
\hline
$\Delta$($M_{\rm f467m}-M_{\rm f814w}$) & 0.194* & 0.007 & $\sim$0.000 & $-$0.028 & $-$0.001 & 0.012 & 0.203* \\
$\Delta$($M_{\rm f487n}-M_{\rm f814w}$) & 0.191* & 0.009 & 0.003 & $-$0.012 & 0.001 & 0.018 & 0.204* \\
$\Delta$($M_{\rm f502n}-M_{\rm f814w}$) & 0.224* & 0.006 & 0.002 & $-$0.137* & 0.002 & 0.110* & 0.260* \\
$\Delta$($M_{\rm f555w}-M_{\rm f814w}$) & 0.153* & 0.006* & $-$0.002 & $-$0.024* & 0.001 & 0.023* & 0.167* \\
$\Delta$($M_{\rm G'}-M_{\rm f814w}$) & 0.124* & 0.006* & $-$0.001 & $-$0.009* & $\sim$0.000 & 0.013* & 0.134* \\
$\Delta$($M_{\rm f606w}-M_{\rm f814w}$) & 0.112* & 0.005* & $-$0.001 & $-$0.011* & $\sim$0.000 & 0.013* & 0.121* \\
$\Delta$($M_{\rm r}-M_{\rm f814w}$) & 0.094* & 0.004 & $-$0.002 & 0.004 & $-$0.001 & $-$0.006* & 0.096* \\
$\Delta$($M_{\rm f656n}-M_{\rm f814w}$) & 0.069 & 0.003 & 0.003 & 0.025 & 0.003 & 0.003 & 0.077 \\
$\Delta$($M_{\rm f658n}-M_{\rm f814w}$) & 0.064* & 0.003 & 0.001 & 0.009 & $\sim$0.000 & $-$0.001 & 0.068* \\
$\Delta$($M_{\rm f673n}-M_{\rm f814w}$) & 0.067* & 0.004 & $\sim$0.000 & 0.004 & $-$0.001 & $-$0.004 & 0.070* \\
\hline
$\Delta$($M_{\rm f815n}-M_{\rm f814w}$) & 0.001 & $-$0.003 & $-$0.003 & $\sim$0.000 & $\sim$0.000 & $-$0.006 & $-$0.006 \\
$\Delta$($M_{\rm f845m}-M_{\rm f814w}$) & $-$0.005 & 0.001 & $\sim$0.000 & 0.001 & $\sim$0.000 & $-$0.006 & $-$0.006 \\
$\Delta$($M_{\rm R'}-M_{\rm f814w}$) & $-$0.003 & $\sim$0.000 & $\sim$0.000 & $\sim$0.000 & $\sim$0.000 & $\sim$0.000 & $-$0.003 \\
$\Delta$($M_{\rm z}-M_{\rm f814w}$) & $-$0.018* & $\sim$0.000 & $\sim$0.000 & $-$0.001 & $\sim$0.000 & 0.001 & $-$0.018 \\
$\Delta$($M_{\rm f925n}-M_{\rm f814w}$) & $-$0.024* & $-$0.008 & $\sim$0.000 & $-$0.001 & $\sim$0.000 & $-$0.004 & $-$0.033* \\
$\Delta$($M_{\rm f850lp}-M_{\rm f814w}$) & $-$0.018* & $-$0.001 & $\sim$0.000 & $\sim$0.000 & $\sim$0.000 & 0.001 & $-$0.018* \\
$\Delta$($M_{\rm y}-M_{\rm f814w}$) & $-$0.026* & $-$0.003 & $\sim$0.000 & $-$0.002 & $\sim$0.000 & $-$0.001 & $-$0.029* \\
$\Delta$($M_{\rm f960m}-M_{\rm f814w}$) & $-$0.026* & $-$0.003 & $\sim$0.000 & $-$0.002 & $\sim$0.000 & $-$0.001 & $-$0.030* \\
$\Delta$($M_{\rm f968n}-M_{\rm f814w}$) & $-$0.024 & $-$0.002 & $\sim$0.000 & $-$0.002 & $\sim$0.000 & $-$0.001 & $-$0.027* \\
\hline 
\end{tabular}
\end{center}
\label{T4}
\end{table*}

\subsection{Carbon, Nitrogen and Oxygen variations}
\begin{figure}
    \centering
    \includegraphics[width=0.9\textwidth]{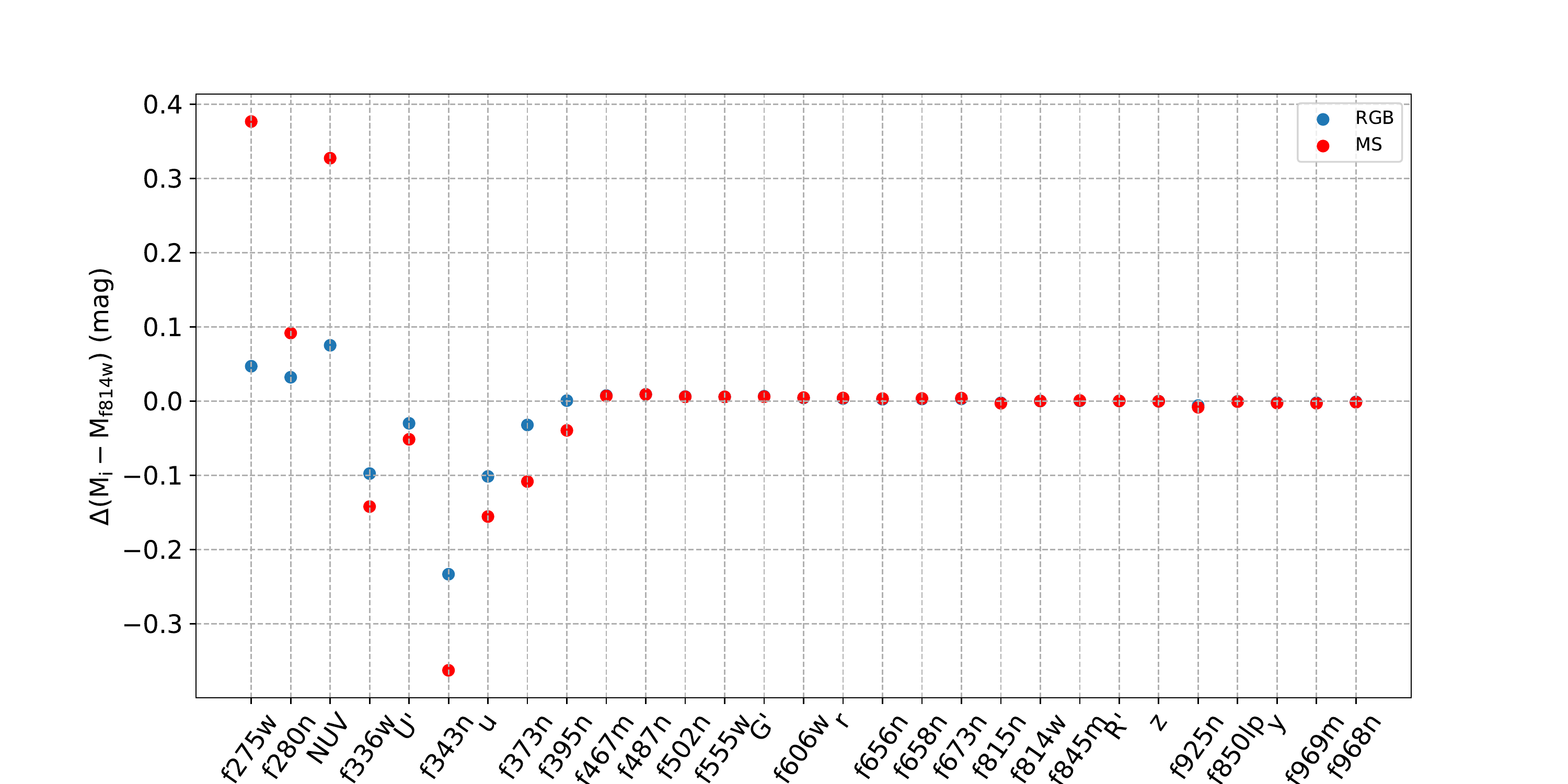}
    \caption{As Fig.\ref{f5}, but for normal and N-enriched (CO-depleted) populations.}
    \label{f6}
\end{figure}

Measuring CNO variations in photometry requires specific filter bands. As introduced in Section~\ref{S1}, filters centered at the wavelength of $\lambda\sim3370\;$\AA (i.e.,CSST-f336w, CSST-f343n) would see N-rich stars fainter than normal stars (at the same stage). Since N-enriched stars are CO-depleted, these stars should be brighter than CO-normal stars in filters covering the wavelength range of $\lambda<3000\;$\AA (OH-absorption dominated) or $\lambda\sim4300\;$\AA (CH-absorption). Unfortunately, the latter is not included in the MCI/{\sl CSST} filter system. We present the color differences between the normal population and the population with CNO anomalies (see Table \ref{T2}) in different color baselines in Fig.\ref{f6}. Indeed, we find that around $\sim3370\;$\AA, normal stars are brighter than N-rich stars. They present negative color differences in color bands involving specific filters (in particular the {\sf CSST-f336w} and {\sf CSST-f343n} filters). For filters bluer than $\sim3000\;$\AA ({\sf CSST-NUV,} {\sf CSST-f280n}, {\sf CSST-f275w}), normal stars are fainter than N-rich stars as the latter are O-depleted, their color differences are positive.

\begin{figure}
    \centering
    \includegraphics[width=0.9\textwidth]{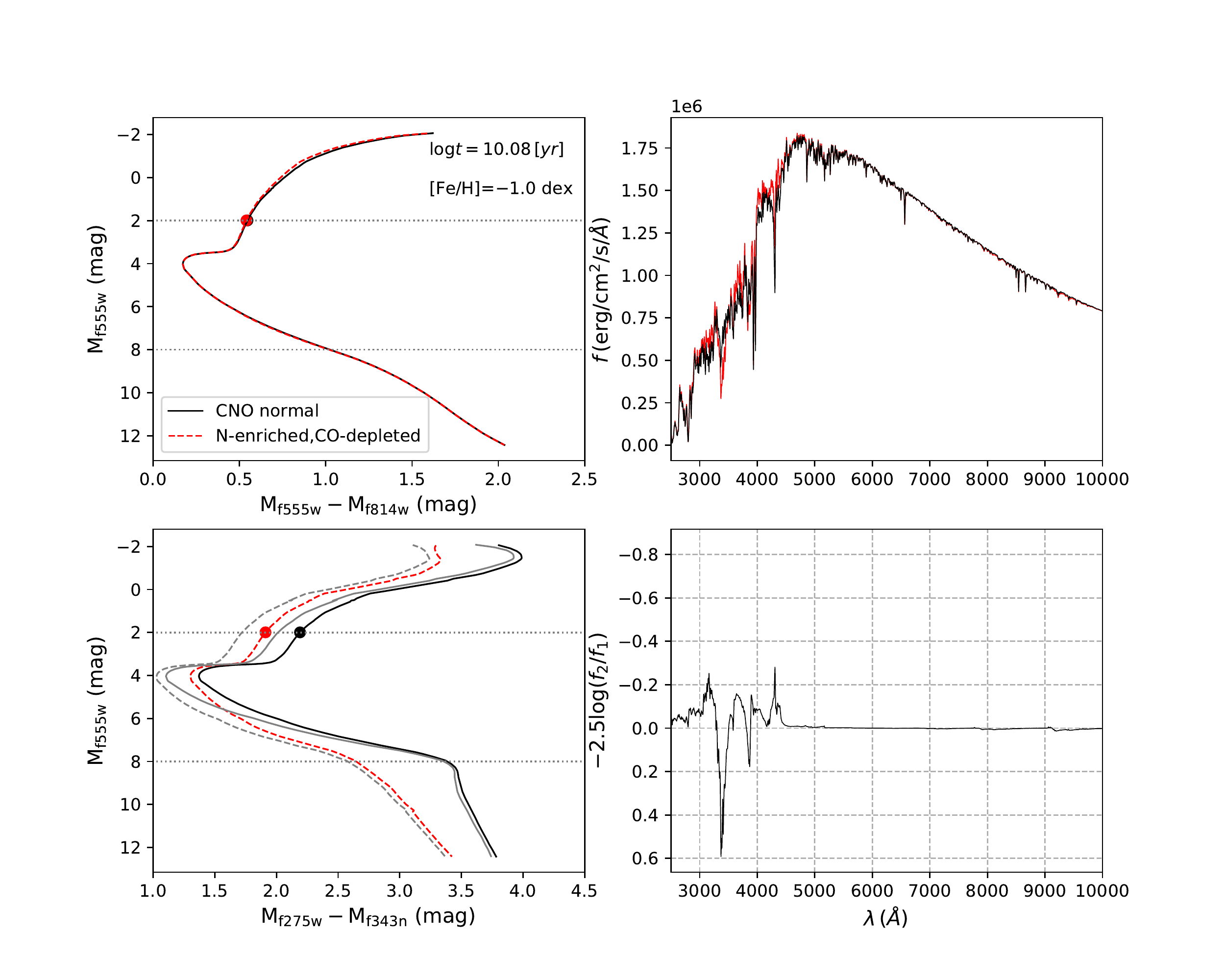}
    \caption{Same as Fig.\ref{f1}, but for normal and N-enriched (CO-depleted) populations (1P and 2P). In left panels we exhibit two loci under colors of $M_{\rm f555w}-M_{\rm f814w}$ (top) and $M_{\rm f275w}-M_{\rm f343n}$ (bottom), respectively. { Grey solid and dashed lines are loci of 1P and 2P in UVIS/WFC3 {\sl HST} filters.} Spectra in right panels are for RGB reference stars.}
    \label{f7}
\end{figure}

Fig.\ref{f6} indicates that if we select a color band involves a filter with $\lambda_{\rm c}<3000\;$\AA ({\sf CSST-f275w}) and a filter with $\lambda_{\rm c}\sim3370\;$\AA ({\sf CSST-f343n}), we can maximize the color difference between the normal and N-rich population stars. It also tells us that the color separation is more significant at the MS stage than the RGB, because the selected MS stage has a surface temperature lower than that of the RGB. Stars with lower atmosphere temperatures will have stronger CNO-related molecular absorptions. In Fig.\ref{f7} we present two CMDs for the normal and N-enriched (CO-depleted) populations (1P and 2P). The upper-left panel presents the CMD involving two visible filter bands, {\sf CSST-f555w} and {\sf CSST-f814w}, which shows a negligible difference between the two populations. The bottom-left panel shows two populations in the color of $M_{\rm f275w}-M_{\rm f343n}$, which exhibits a dramatic color difference. Indeed, similar filter bands are frequently used in {\sl HST} photometry for  studying MPs of GCs \citep[e.g.,][]{Milo18a,Nard18a}. { We present these loci under the same UVIS/WFC3 {\sl HST} filters in the panel. Again, the color differences between the 1P and 2P revealed by {\sl CSST} and {\sl HST} are similar.} In the right panels of Fig.\ref{f7}, we present spectra of the referenced RGB star and its counterpart with CNO anomalies (upper) and their magnitude difference spectrum (bottom), which explains why the combination of {\sf CSST-f275w} and {\sf CSST-f343n} can maximize their color difference. The color differences caused by CNO variation in different filter bands are present in Tables \ref{T3} and \ref{T4}. 

\subsection{Sodium variation}
\begin{figure}
    \centering
    \includegraphics[width=0.9\textwidth]{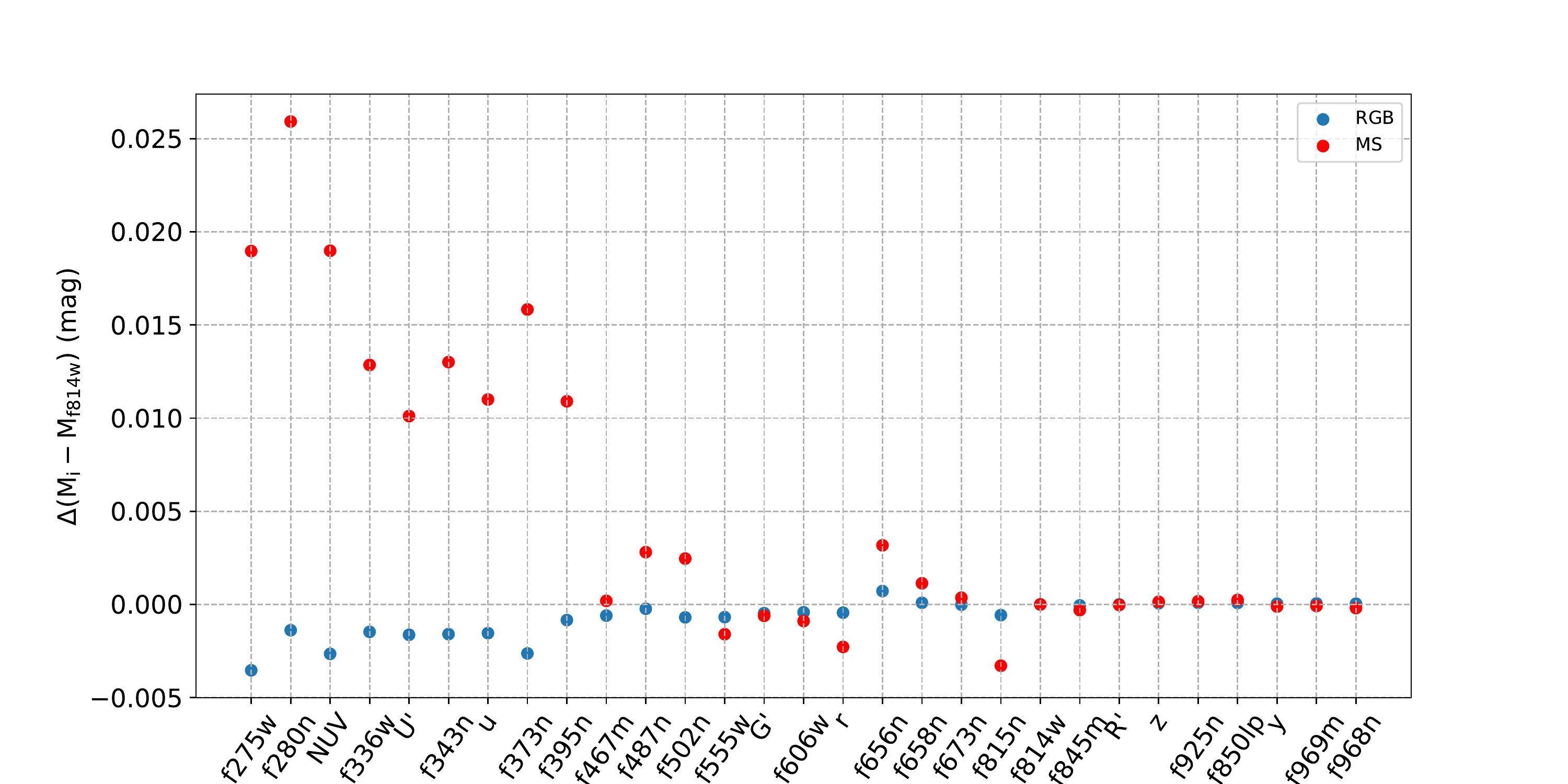}
    \caption{As Fig.\ref{f5}, but for normal and Na-enriched populations.}
    \label{f8}
\end{figure}

We apply the same analysis to stellar populations with different Na abundance. Although for MPs the Na and O abundances are always anti-correlated, in this subsection we only consider a Na enrichment of $\delta{\rm [Na/Fe]}$=0.5 dex. More realistic considerations (NGC 2808-like) will be discussed later. Our analysis shows that Na enrichment has very limited effect on photometry. Although lots of Na I and Na II absorption lines occur in UV band, the maximum color difference between the two populations does not exceed 0.03 mag. We also find that Na-enrichment has opposite effects on the MS and RGB stages. Na-rich stars are slightly bluer than normal stars at the RGB stage, but are redder at the bottom MS, as shown in Fig.\ref{f8}. We conclude that the filters designed for MCI/{\sl CSST} are not suitable to identify MPs with different Na abundance. 

\subsection{Magnesium variation}
\begin{figure}
    \centering
    \includegraphics[width=0.9\textwidth]{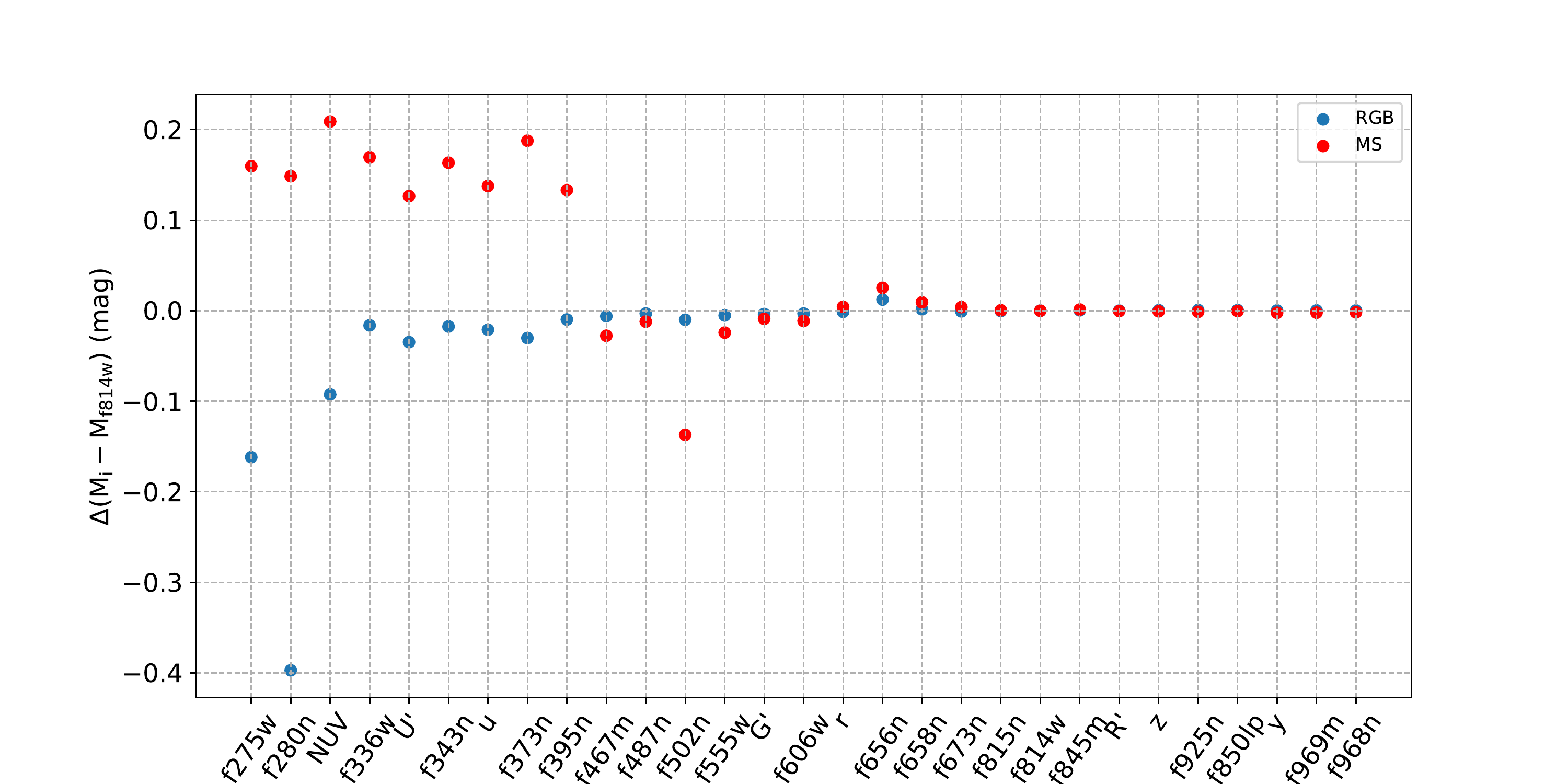}
    \caption{Same as Fig.\ref{f5}, but for normal and Mg-enriched populations.}
    \label{f9}
\end{figure}

The most significant difference between the normal and Mg-rich population stars appears in the color band of $M_{\rm f280n}-M_{\rm f814w}$. Our analysis shows that for the RGB stage, their color difference reaches $\Delta{(M_{\rm f280n}-M_{\rm f814w})}=-$0.4 dex. This is caused by the combination of  the MgII 2795\AA\;and 2803\AA\, absorption lines. Intriguingly enough, the color difference is reversed for the bottom MS populations. The Mg-rich population is bluer than the normal population in most color bands involving UV filters. but is redder in $M_{\rm f502n}-M_{\rm f814w}$. Since its chemical abundances is identical to the RGB, this difference is possibly caused by their surface gravity ($\log{g}$) difference. Since the degree of ionization depends on the stellar surface gravity, its mainly affects the wavelength range below 3650\;\AA. Indeed, as shown in Fig.\ref{f9}, filters with central wavelength below 4000\;\AA\, are significantly affected. The lower surface temperature may also affect the SED: for very late stars, the enhanced Mg abundance will produce a deeper absorption band includes Mg I 5167\AA,  5173\AA\; and 5184\AA\; than normal star, which thus affects the {\sf CSST-f502n} photometry. Since to explain the details of this difference is beyond the scope of this article, we leave it an open question for future investigation. 

\begin{figure}
    \centering
    \includegraphics[width=0.9\textwidth]{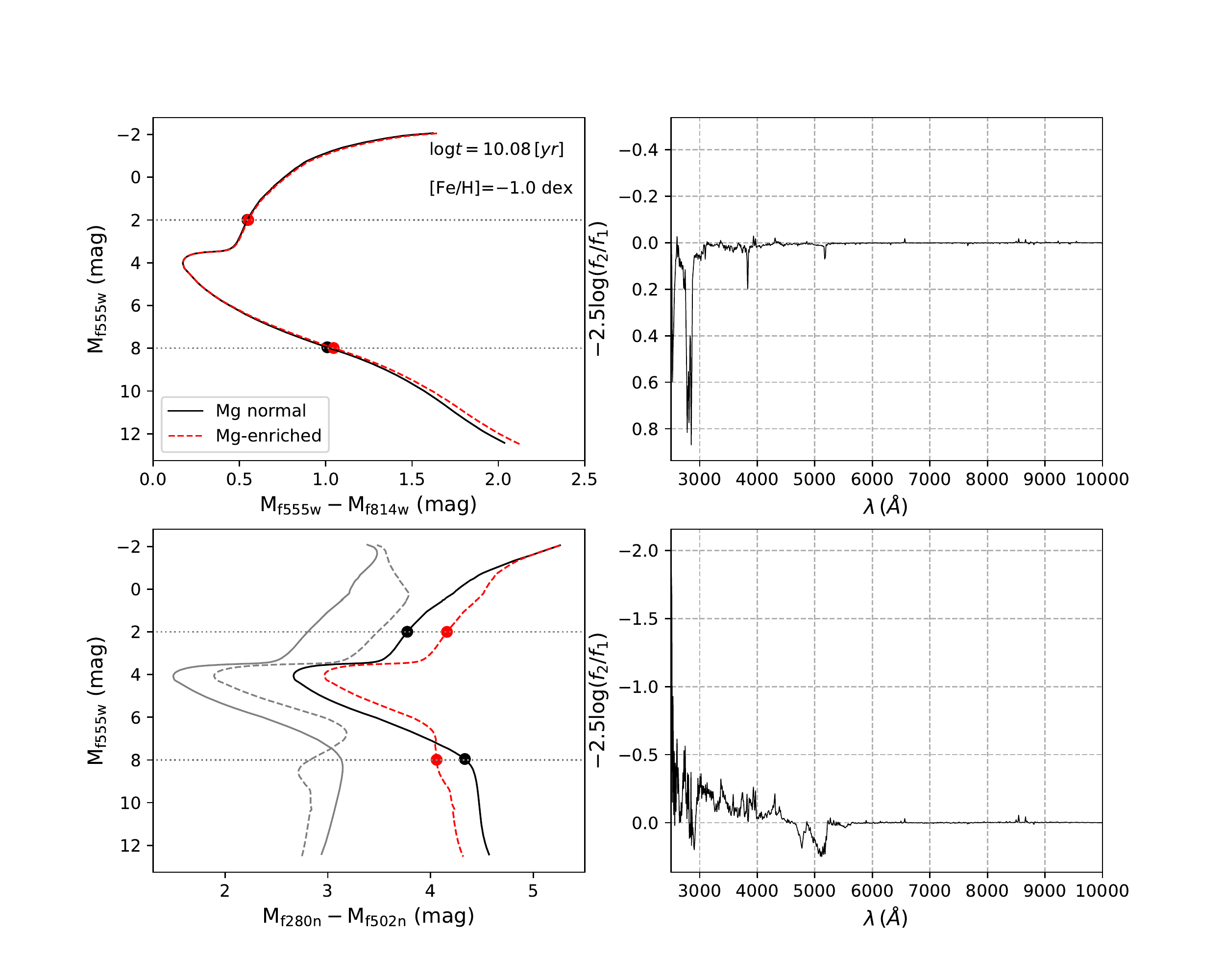}
    \caption{Same as Fig.\ref{f7}, but for normal and Mg-rich populations. In right panels we present the magnitude difference spectra between normal and Mg-rich RGB (top) and MS (bottom) stars ($f_2$ and $f_1$ are fluxes of 2P and 1P stars).}
    \label{f10}
\end{figure}

Fig.\ref{f9} indicates that the separation between the normal and Mg-rich is most significant in the color band of $M_{\rm f280n}-M_{\rm f502n}$. We present the CMDs of normal (1P) and Mg-rich (2P) populations in the left panels of Fig.\ref{f10}. { We find the optimal filter sets used for separating the normal and Mg-rich populations are {\sf CSST-f280n} and {\sf CSST-f502n} (for {\sl HST}, these are {\sl F280N} and {\sl F502N} accordingly). In these two filters, the color difference between 1P and 2P are much more significant than that observed in optical passbands (i.e., $M_{\rm f555w}-M_{\rm f814w}$). The color difference seen by the {\sl CSST} is less significant than that observed by {\sl HST} at MS, however. Because the transmission curve of {\sl HST} F280N filter is higher than that of {\sl CSST}, thus the former is more sensitive to small Mg variation.} In the right panels of Fig.\ref{f10} we present the magnitude difference spectra between the Mg-rich and normal referenced stars (top-right: RGB stars; bottom-right: bottom MS stars). 

\subsection{Aluminum variation}
\begin{figure}
    \centering
    \includegraphics[width=0.9\textwidth]{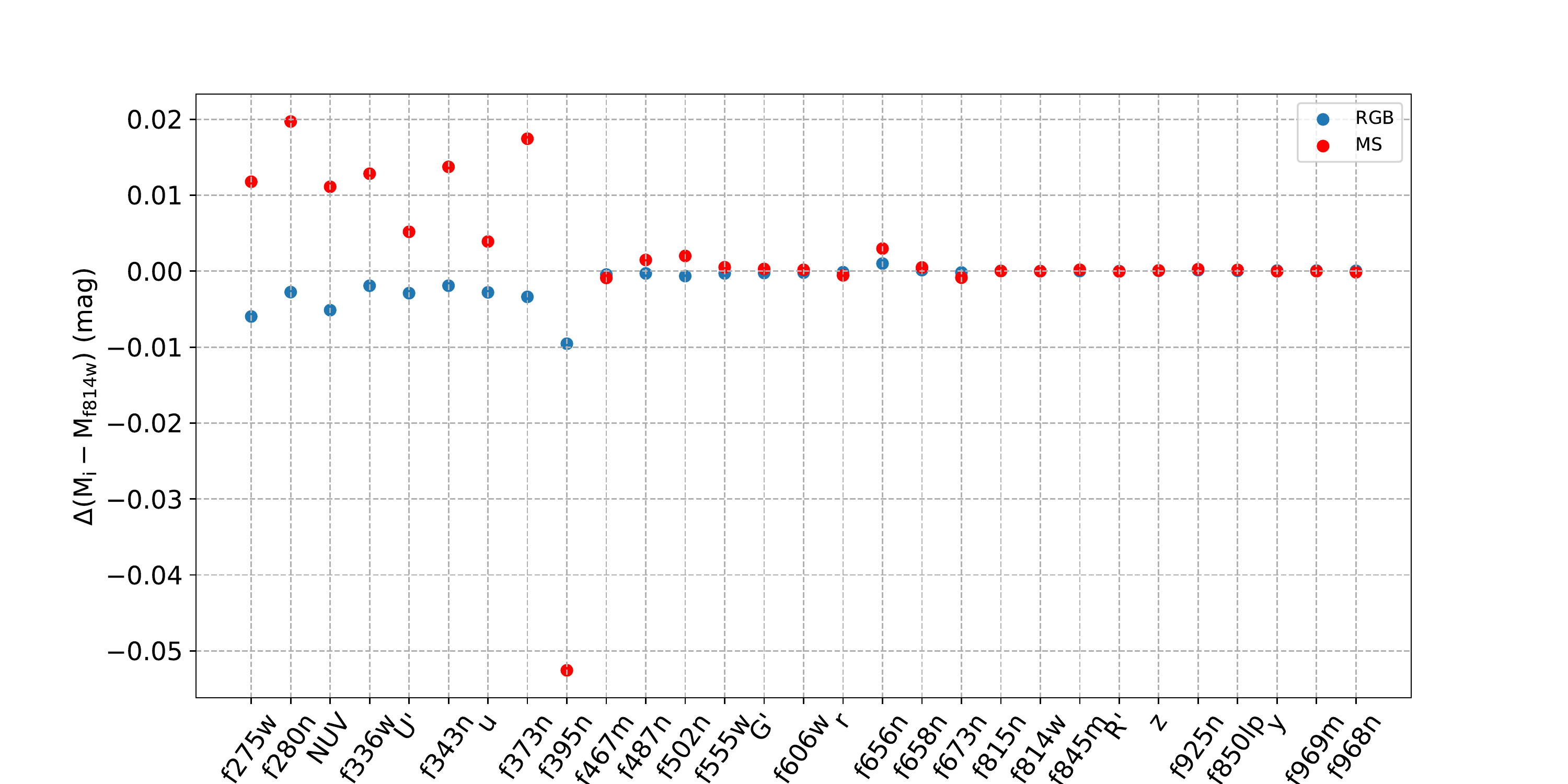}
    \caption{As Fig.\ref{f5}, but for normal and Al-rich populations.}
    \label{f11}
\end{figure}

The effect of Al-enrichment is similar to that of Mg (Fig.\ref{f11}). When considering the colors of $M_{\rm i}-M_{\rm f814w}$, where $M_{\rm i}$ only contains filter bands bluer than 3800\;\AA, the Al-rich population is redder than the normal population at RGB stage, but is bluer at the bottom MS range. These color differences are very small ($|\Delta(M_{\rm i}-M_{\rm f814w})|<0.02$ mag). Only for the bottom MS range in the color band of $M_{\rm f395n}-M_{\rm f814w}$, the Mg-rich population is redder than the normal population with $\Delta(M_{\rm i}-M_{\rm f814w})<-0.05$ mag. This is caused by the combination of Al I lines at 3944\,\AA\, and 3962\,\AA. However, resolving such a small color difference at the bottom MS in a narrow filter band may require a very long exposure time (see Section \ref{S4}). We, therefore, do not recommend using the {\sl CSST} photometry to study Al variations in star clusters. 

\subsection{NGC 2808-like multiple populations}
Above we only discuss the effects of MPs with variations of a few or individual elements. A more realistic case should involve variations in all these elements. In this work, we only consider the variations of He, C, N, O, Na, Mg and Al because they are the most abundant elements in stellar atmospheres, producing strong absorption features. For GCs with MPs, Li, F, and some s-process elements may vary from star-to-star as well. Because of their small mass fraction, strong photometric effects caused by these variations are not expected. In this section, we study two cases that follow the definitions in Table \ref{T2}. The difference between these two cases is that Case 1 does not contain He variation, because not all GCs have such an extreme He variation like NGC 2808. 

\begin{figure}
    \centering
    \includegraphics[width=0.9\textwidth]{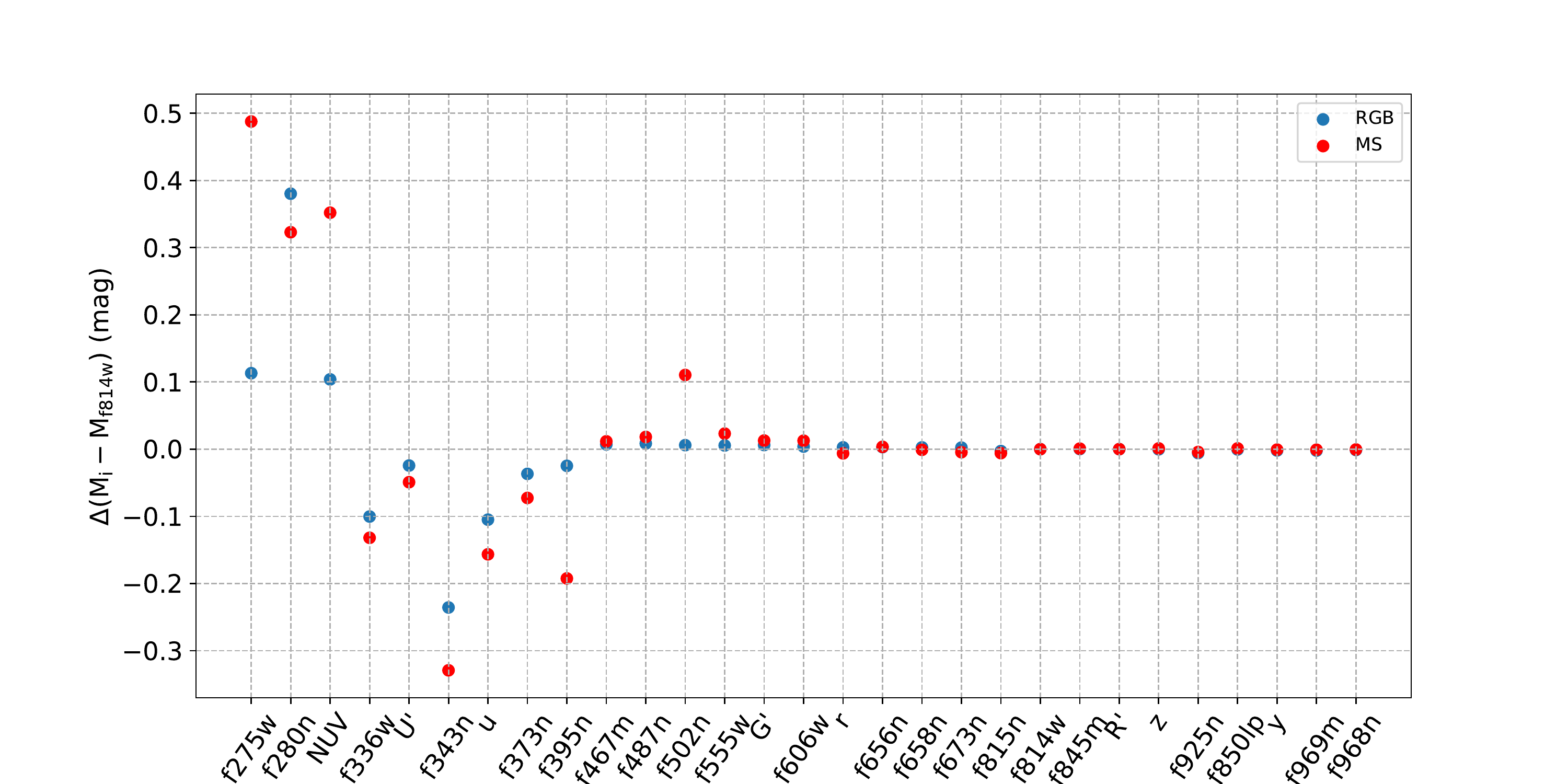}
    \caption{As Fig.\ref{f5}, in this figure, the 2P have variations in both the C, N, O, Na, Mg, Al elements, as defined in Table.\ref{T2} (Case 1)}
    \label{f12}
\end{figure}

\begin{figure}
    \centering
    \includegraphics[width=0.9\textwidth]{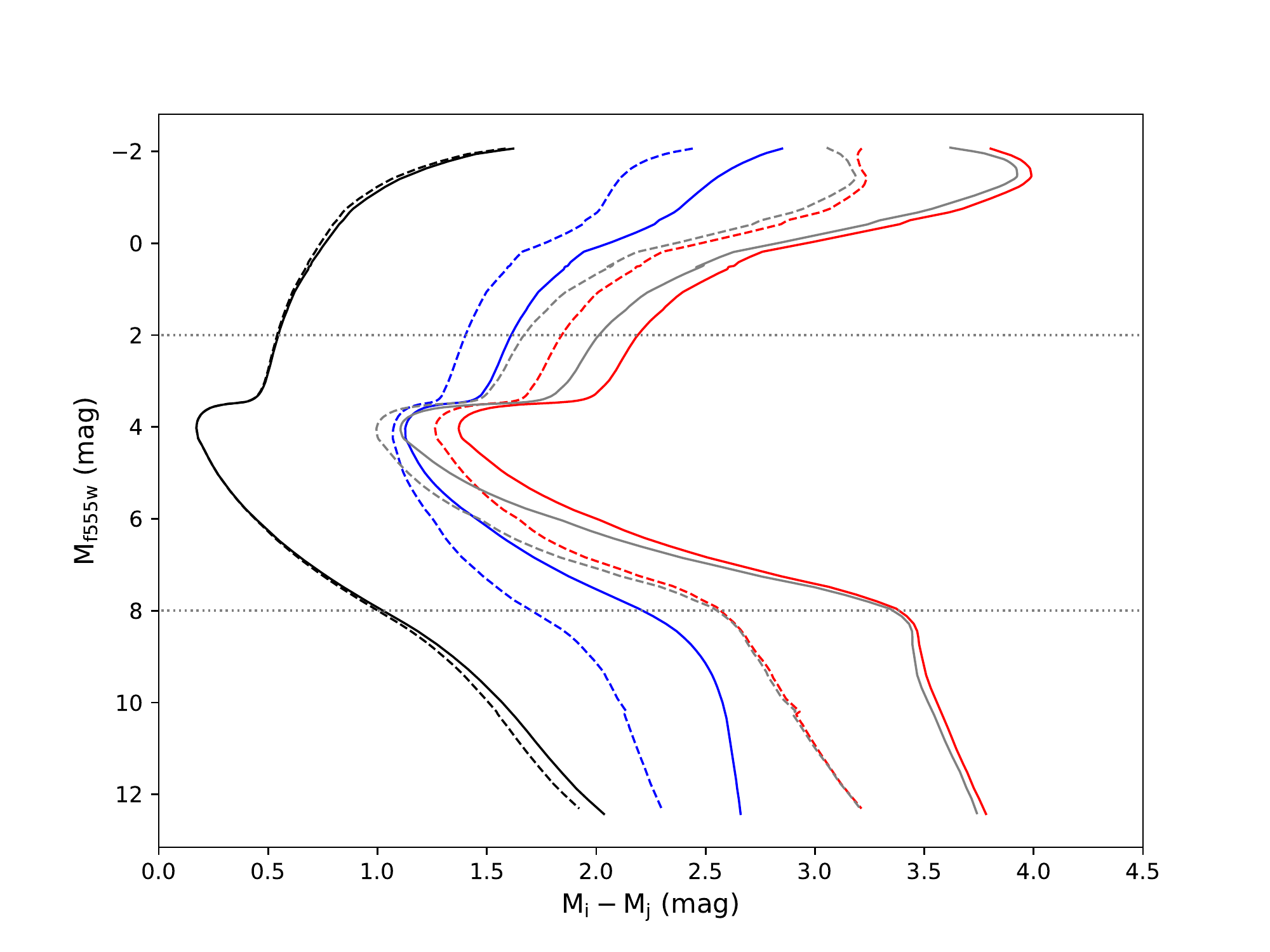}
    \caption{Loci of 1P and 2P (Case 1) in different color bands (black: $M_{\rm f555w}-M_{\rm f814w}$; blue: $M_{\rm NUV}-M_{\rm u}$; red: $M_{\rm f275w}-M_{\rm f343n}$; { grey: $M_{\rm F275W}-M_{\rm F343N}$, for {\sl HST}}). The solid and dashed lines represent 1P and 2P stars.}
    \label{f13}
\end{figure}

The color differences between 1P and 2P for case 1, are shown in Fig.\ref{f12}. Not surprisingly, UV filter bands play important roles in revealing MPs. We find that the color band of $M_{\rm f275w}-M_{\rm f343n}$ can maximize the difference between two populations. We also suggest an alternative color band of $M_{\rm NUV}-M_{\rm u}$. Although the color difference between the 1P and 2P is not so significant in this band like $M_{\rm f275w}-M_{\rm f343n}$, the {\sf CSST-NUV} and {\sf CSST-u} filters have much wider FWHM than the {\sf CSST-f343n} (see Fig.\ref{f1} and Fig.\ref{f2}), which can save lots of exposure time. In addition, these two filter bands will be used in the CSST main survey as well \citep{Gong19a}. In Fig.\ref{f13} we show the CMDs of two populations in CMDs involving colors of $M_{\rm f275w}-M_{\rm f343n}$, $M_{\rm NUV}-M_{\rm u}$ and $M_{\rm f555w}-M_{\rm f814w}$. { The isochrones described by {\sf F275W} and {\sf F343N} passbands of UVIS/WFC3 {\sl HST} are present by grey solid and dashed lines, respectively. The {\sl CSST} and {\sl HST} can well separate the loci for 1P and 2P, and their separations are similar. }

\begin{figure}
    \centering
    \includegraphics[width=0.9\textwidth]{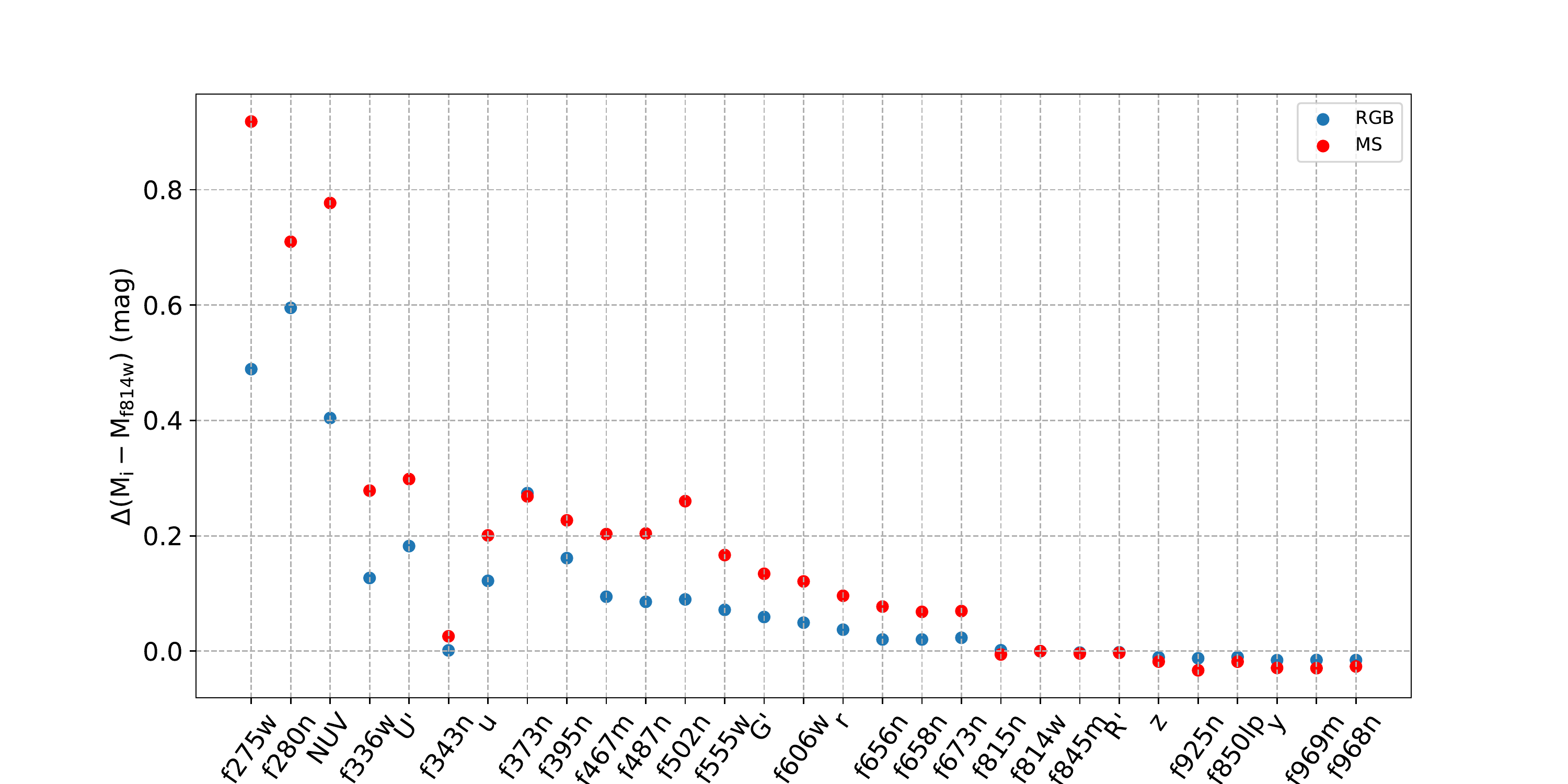}
    \caption{As Fig.\ref{f5}, in this figure, the 2P have variations in both the He, C, N, O, Na, Mg, Al elements, as defined in Table.\ref{T2} (Case 2)}
    \label{f14}
\end{figure}

\begin{figure}
    \centering
    \includegraphics[width=0.9\textwidth]{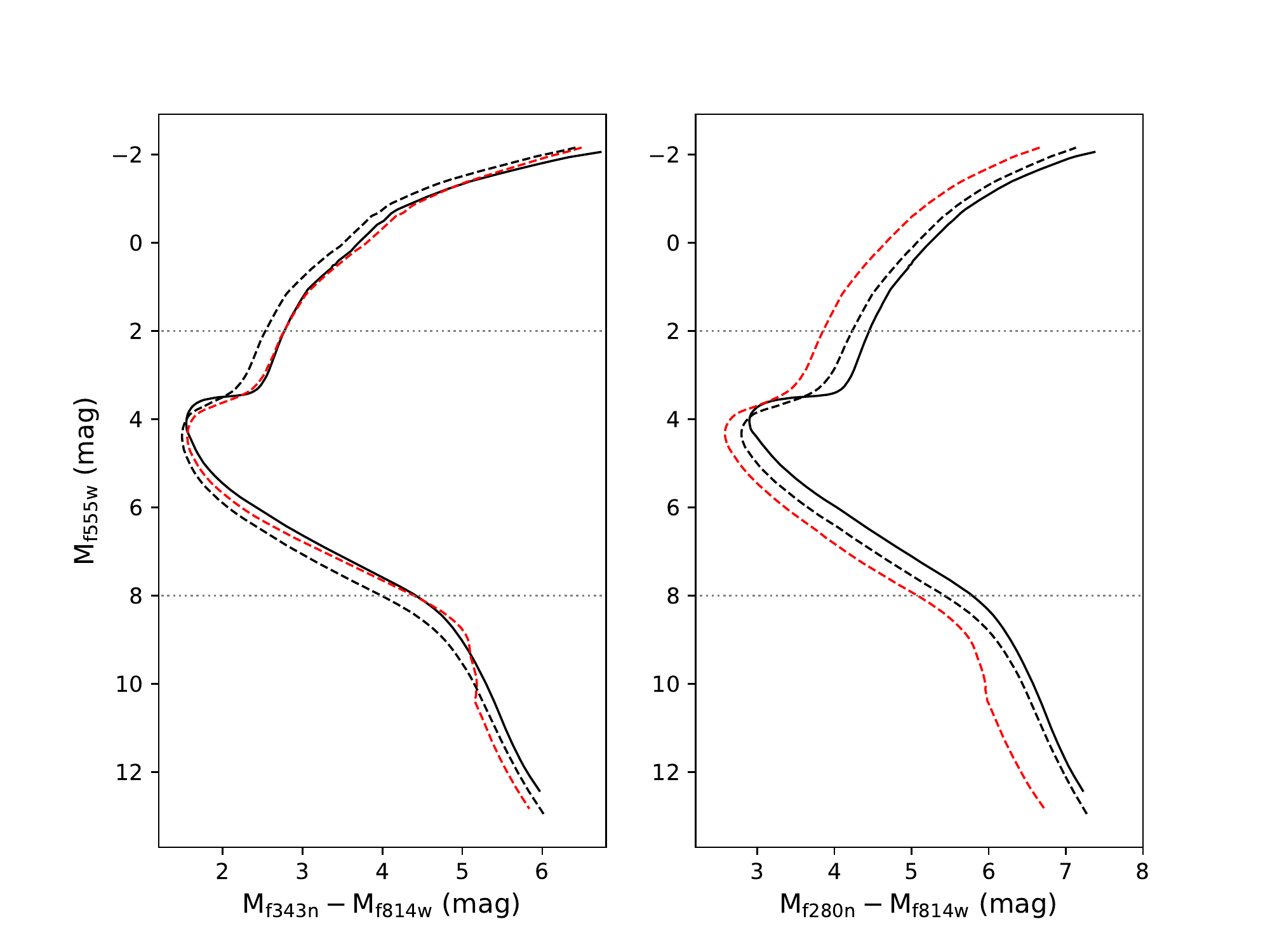}
    \caption{CMDs of 1P (black solid line), 2P (Case 2, red dashed line) and He-rich population (without other element variations, black dashed line), in the color bands of $M_{\rm f343n}-M_{\rm f814w}$ (left) and $M_{\rm f280n}-M_{\rm f814w}$ (right).}
    \label{f15}
\end{figure}

The color differences between 1P and 2P stars for Case 2 are present in Fig.\ref{f14}. Driven by the He-enrichment, 2P becomes bluer in all color bands. For this case, the color difference between the 1P and 2P bottom MS can reach $\Delta(M_{\rm f275w}-M_{\rm f814w})>0.8$ mag. Although He variation produces the most significant color difference, we can derive other element variations through different color bands. For example, in Fig.\ref{f5} we can see if the 1P and 2P are only different in He abundance, they will show a color difference of $\Delta(M_{\rm 343n}-M_{\rm f814w})\sim$0.2--0.5 mag. This difference becomes negligible if they are also different in CNO abundances, because N-enrichment (CO-depletion) will make the 2P stars redder than normal stars, compensating the effect of He-enrichment. In addition, if the He-rich population is Mg-depleted, they will become much bluer in $M_{\rm f280n}-M_{\rm f814w}$ color band ($\Delta(M_{\rm f280n}-M_{\rm f814w})\sim$0.6--0.8 mag) than the case without Mg-depletion ($\Delta(M_{\rm f280n}-M_{\rm f814w})\sim$0.2--0.4 mag, Fig.\ref{f5}). These are illustrated in Fig.\ref{f15}. 

\subsection{A less extreme case}
\begin{figure}
    \centering
    \includegraphics[width=0.9\textwidth]{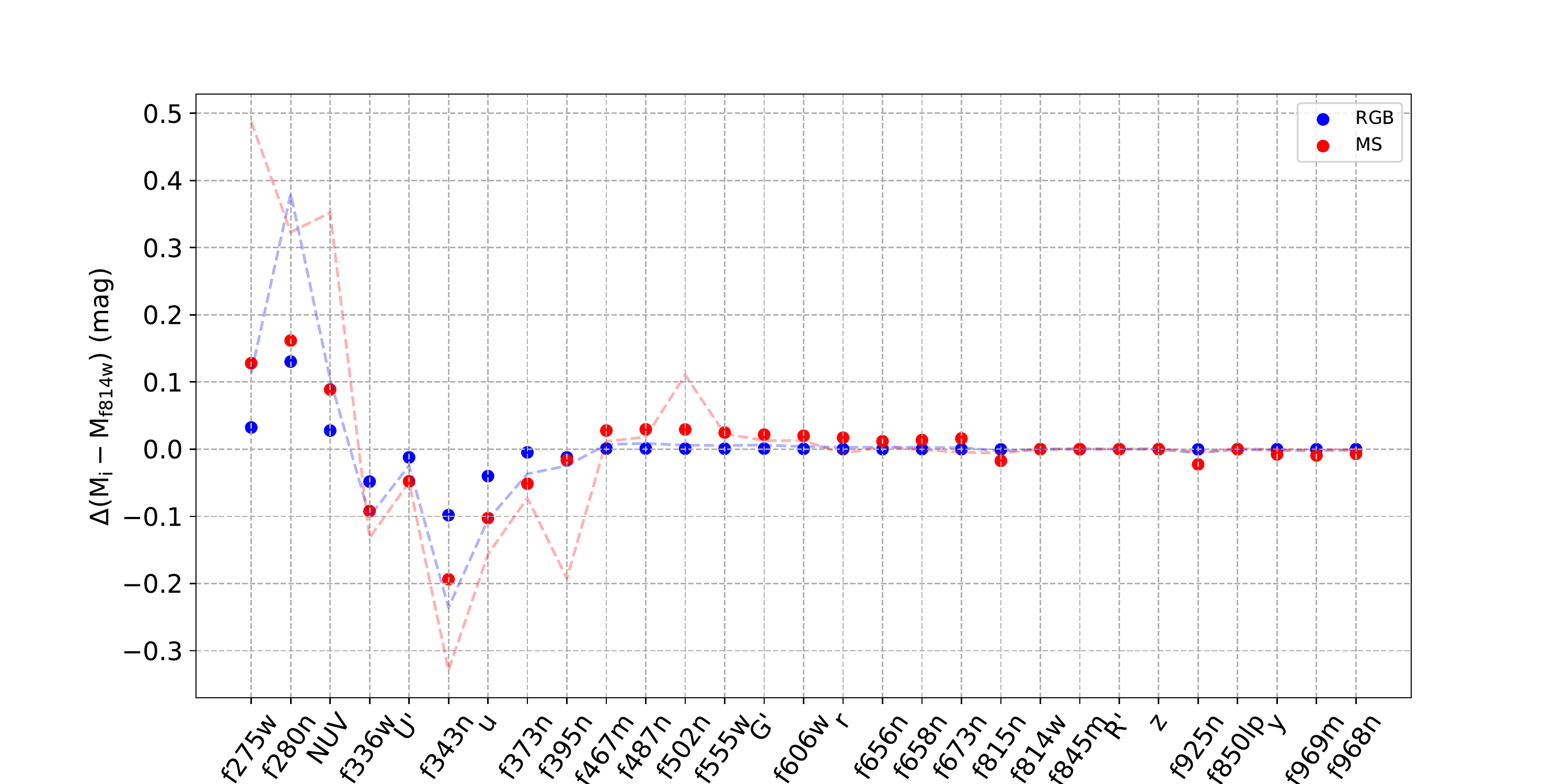}
    \caption{As Fig.\ref{f5}, but for a less extreme case (Case 3), as defined in Table.\ref{T2}. As a comparison, the Case 1 color differences are plotted by red (MS) and blue (RGB) dashed lines.}
    \label{f16}
\end{figure}

{ We have calculated a less extreme model of MPs. According to \cite{Milo18a}, most less massive GCs ($\lesssim10^5$ $M_{\odot}$) do not exhibit a significant helium spread ($\delta{Y}\sim0.00-0.02$, their figure 13). But these clusters still exhibit light element variations (their table 3). Most GCs have N, Mg, and Al variations that are about half of the NGC 2808. In this case, we set an N-enrichment for 2P stars of $\Delta$[N/Fe]=0.7 dex, and these 2P stars are depleted by $\Delta$[C/Fe]=$\Delta$[O/Fe]=$-$0.18 dex, to make the total CNO abundance constant. Their Mg-depletion is $\Delta$[Mg/Fe]$=-0.2$ dex, with an Al-enrichment of $\Delta$[Al/Fe]$=+0.7$ dex, determined through vision inspection from \cite{Panc17a}. The Na-enrichment is $\Delta$[Na/Fe]$=+0.2$ dex. In this case, the N-enrichment and Mg-depletion for 2P stars are half of that in NGC 2808. Using the same method, we have calculated the color differences between 1P and 2P for RGB and MS stars, which are present in Fig.\ref{f16}. We find that under this case, the color differences of $M_{\rm i}-M_{\rm f814w}$ ($M_{\rm i}$ is the magnitude under any selected filter) are less obvious than in Case 1, as expected. But they still produce significant color differences when involving UV filter bands. The $M_{\rm f343n}$ is the most important filter band for separating 1P and 2P, which describes the depth of the NH-absorption feature. $M_{\rm f275w}$, $M_{\rm f280n}$ and $M_{\rm NUV}$ can maximize the color difference as they measure the O-depletion. The color band like $M_{\rm f343n}-M_{\rm f275w}$ is more sensitive than traditional optical colors (e.g., $V-I$).}

In summary, multi-bands photometry of the MCI/{\sl CSST} involving these key filters is crucial for determining detail abundance variations between different stellar populations. 

\section{Discussion}\label{S4}

In this work we study the photometric patterns of MP chemical variations on synthetic magnitudes of the MCI/{\sl CSST} filter system. We studied five cases,  including He variation, CO-depletion and N-enrichment, Na-enrichment, Mg-depletion and Al-enrichment, two cases with different chemical patterns similar to the GC NGC 2808 (with/without He variation), { and one less extreme case which better represents many other GCs.} We find that colors involving various UV filters are well suited to separating MPs with different He, C, N, O, Mg abundances, but are not suitable for Na and Al variations. We find that the filter {\sf CSST-f343n} is essential for deriving CNO variations. The color band involving the {\sf CSST-f280n} filter is optimal for separating MPs with different Mg abundance. { The performances of these filters are similar to their counterparts in UVIS/WFC3 {\sl HST} photometric system.} Considering the exposure time requirement, we suggest that wide filter bands such as {\sf CSST-f275w}, {\sf CSST-NUV}, and {\sf CSST-u} can be used for studying MPs in star clusters. 


Although currently simulated artificial images for MCI/{\sl CSST} observations are not available, it is useful to have a preliminary evaluation of whether one can disentangle MPs with MCI/{\sl CSST} at a typical distance of GCs. We make use of the online {\sl CSST} exposure time calculator (ETC) for the MCI\footnote{http://etc.csst-sc.cn/ETC-nao/etc.jsp}, to study if we can resolve the referenced RGB and bottom MS stars at the distance of NGC 2808 ($(m-M)_0\sim15.6$ mag, \cite{Kund13a}) in given color bands. For two RGB stars ($m_{\rm f555w}=M_{\rm f555w}+15.6=$17.6 mag), we adopt one exposure, with a total exposure time of $\sim$300 s, while for the bottom MS stars ($m_{\rm f555w}=M_{\rm f555w}+15.6=$23.6 mag), we set totally 18 exposures with an exposure time of $300\times180=$54,000 s (With this exposure time, the SNR for {\sf CSST-f814w} at the bottom MS is roughly the same to the SNR at the RGB phase with 300 s exposure time). For a selected filter, the ETC will return its SNR under a given exposure time. If the resulting SNR is four times the minimum requirement for disentangling MPs, we remark the filter as a suitable filter (denoted by asterisks in Tables \ref{T3} and \ref{T4}). We find that using MCI/{\sl CSST} to resolve NGC 2808-like MPs at the bottom MS is feasible through specific color bands (i.e., $M_{\rm NUV}-M_{\rm f814w}$). 

Of course, the real {\sl CSST} observations will be definitely somehow different from what we have evaluated in this work. For example, although Tables \ref{T2} and \ref{T3} report that the color band of $M_{\rm f555w}-M_{\rm f814w}$ can disentangle MPs with different CNO abundances, their color differences are only $\Delta(M_{\rm f555w}-M_{\rm f814w})\sim$0.006 mag. Such a small color difference can be easily contaminated by unresolved binaries, blending, differential reddening and point spread function (PSF) fitting residuals. Because of this, a suitable filter does not mean an optimal selection for studying MPs with certain chemical patterns. Anyhow, our analysis definitely shows that MCI/{\sl CSST} will be a powerful tool for studying MPs in GCs. This work can be served as guidance for arranging future MCI/{\sl CSST} observations, such as the choice of filter sets and benchmark GCs. 

\begin{acknowledgements}
This work was supported by National Natural Science Foundation of China (NSFC) (Grant Nos. 12073090), and the China Manned Space Project with NO.CMS-CSST-2021-A08, CMS-CSST-2021-B03. We thank Dr. Licai Deng for expertly commented the paper. We thank Dr. Yang Chen for calculating bolometric corrections.
\end{acknowledgements}

\bibliographystyle{raa}
\bibliography{my_bib}

\label{lastpage}
\end{CJK*}
\end{document}